\documentclass[aps, pra, a4paper, showpacs, twocolumn, english, 10pt]{revtex4-1}
\usepackage{bbm, amsthm, bm, textcomp, nicefrac, geometry, ragged2e}
\usepackage[dvipsnames]{xcolor}
\usepackage{float}
\usepackage[bbgreekl]{mathbbol}
\usepackage{graphicx, epstopdf, color, verbatim, enumitem, ulem}
\usepackage{pbox,array}
\usepackage{mathrsfs}
\usepackage{amssymb}
\usepackage{lipsum,ulem}
\usepackage{textgreek}
\usepackage{mathtools}
\usepackage{stackrel}
\usepackage[thinlines]{easytable}
\usepackage{amsmath}
\usepackage[utf8]{inputenc}
\makeatletter
\usepackage{soul}
\usepackage[caption=false]{subfig}
\usepackage{hyperref}
\hypersetup{
    colorlinks = true,
    linkcolor = Purple,
    citecolor = Purple,
    filecolor = Purple,
    urlcolor = Purple}
\usepackage[T1]{fontenc}
\usepackage[caption=false]{subfig}
\usepackage{babel}
\usepackage[linesnumbered,ruled]{algorithm2e}

\addto\captionsenglish{}

\newcommand\id{\mathbbm{1}}

\newcommand{\ket}[1]{\left| #1 \right\rangle}
\newcommand{\bra}[1]{\left\langle #1 \right|}
\newcommand{\ketbra}[2]{\left| #1\right\rangle\!\left\langle#2\right|}

\newcommand{\be}{\begin{equation}}
\newcommand{\ee}{\end{equation}}
\newcommand{\bea}{\begin{eqnarray}}
\newcommand{\eea}{\end{eqnarray}}

\definecolor{brickred}{rgb}{0.8, 0.0, 0.0}

\begin{document}

\title{How many lives does Schrödinger's cat have?}

\author{Andrea L\'opez-Incera$^1$, Wolfgang D\"ur$^1$ and Stefan Heusler$^2$}

\affiliation{$^1$Universit\"at Innsbruck, Institut f\"ur Theoretische Physik, Technikerstra{\ss}e 21a, 6020 Innsbruck, Austria \\ $^2$Institut für Didaktik der Physik, Wilhelm-Klemm-Str. 10, 48149 Münster, Germany}

\date{\today}

\begin{abstract}
Schr\"odinger's cat is an iconic example for the problem of the transition from the microscopic quantum world to the macroscopic, classical one. It opened many interesting questions such as, could a macroscopic superposition like a dead and alive cat ever exist? What would be the characteristic features of such a system? The field of macroscopicity aims at providing answers to those questions, both from a theoretical and an experimental point of view. Here, we present the main concepts in macroscopicity, including macroscopicity measures, experimental realizations and the link to metrology, from a pedagogical perspective. We provide visualizations and intuitive explanations, together with a hands-on activity where students can create their own macroscopic quantum cats from cardboard cells that are in a superposition of being dead and alive. 
\end{abstract}

\maketitle

\section{Introduction}

In spite of the fact that Schr\"odinger's cat has been dead from the very beginning, it seems that it will never die in the history of physics: Schr\"odinger himself never assumed that a superposition of  a ``dead'' and an ``alive'' cat could ever exist literally. Metaphorically, he just wanted to point to a very fundamental problem in quantum physics: How to describe the \textit{transition} between the quantum and the classical, macroscopic world? It is this question that persists to be notoriously difficult even 100 years after the birth of quantum physics. 

 Schr\"odinger's cat is an ordinary, macroscopic cat, like the one you might have as a pet, made up of millions of atoms. The cat is placed---alive---inside a chamber that is then sealed. Inside the chamber, there is an atom, i.e., a microscopic quantum system, coupled to a mechanism that, when activated, releases a poison that kills the cat. The precise activation mechanism is not important here; what matters is that the poison is released only when the quantum system is in a decayed state. According to quantum mechanics, such a quantum system, prior to any measurement, is described as being in a superposition of decayed and non-decayed states. This leads to the conclusion that the cat is also in a superposition of being both dead and alive, and the total system in the box is described as
\begin{equation*}
    \ket{\textrm{decayed atom, dead cat}} + \ket{\textrm{non-decayed atom, alive cat}}. 
\end{equation*}
This last conclusion was used by Schr\"odinger to highlight one of the flaws of the Copenhagen interpretation of quantum mechanics, because such a macroscopic superposition of a cat being both dead and alive cannot exist. 

Apart from initiating discussions about the interpretations of quantum mechanics, this thought experiment inspired many other research lines over the years. One of them is nowadays termed \textit{macroscopicity} and is the topic of this manuscript.

Instead of deeming the superposition of a dead and alive cat impossible, scientists set out to find ways to create such macroscopic superpositions and started to study the properties that these quantum systems would have if they existed. As years went by, they discovered other systems that are macroscopic in some sense and show effects that can only be explained by quantum mechanics. One example is the specific heat of insulators, which shows a $T^3$ dependence on the temperature $T$ in the regime of small temperatures that can only be predicted by the Debye model, based on quantum mechanics. Another example is superconductivity,  which is microscopically described by entangled pairs of electrons known as Cooper pairs. Both insulators and superconductors are materials that exhibit quantum effects in the macroscopic world—--effects that can be observed with the naked eye, such as a bulk of superconducting material levitating above a magnet. Having these different examples where quantum mechanics is at the core of the dynamics of a macroscopic system inevitably led to questions like, what is the difference between a Schr\"odinger-cat type of system and a large system that shows accumulated, microscopic quantum effects? The field of macroscopicity is devoted to answering these questions and to deeply understand what we will call ``macroscopic quantum states'' in the following, both from an  experimental and a theoretical perspective.

In this article, we review the present state of the art in this field from a pedagogical perspectice. Moreover, we provide simple visualizations in order to support an intuitive approach to the rather abstract mathematical concepts important in the field. We complement the visualizations with a hands-on activity in which quantum cats can be created from cells made out of cardboard and later analyzed from the point of view of macroscopicity. Finally, we discuss relevant experiments and actual applications in the field of quantum technology, in particular, in quantum metrology. 

The article is organized as follows. We present the main ideas and intuitions on macroscopicity in Section~\ref{sec:effective_size}, where we also introduce the main characters of these manuscript: two cats, A and B, that are, respectively, a macroscopic quantum state and a quantum state that is large but not macroscopically quantum. In Section~\ref{sec:macroscopicity_measures}, we use our cats to explain some of the most relevant macroscopicity measures in a pedagogical, visual way. In Section~\ref{sec:transition}, we introduce more complex and realistic states that lie in between our cat A and our cat B, and their experimental realizations. In Section~\ref{sec:hands-on-activity}, we propose a hands-on activity to physically create quantum cats with cardboard cells and visualize and study their macroscopicity properties. We establish the connection between the fields of macroscopicity and metrology in Section~\ref{sec:macrostates_and_metrology} and finish the article with Section~\ref{sec:DeadAlive}, where we present some state-of-the-art experiments that created macroscopic quantum states.

\section{Macroscopicity: initial ideas and effective size}\label{sec:effective_size}

Let us start with the notion of macroscopic or ``large''. Having the picture of the cat in mind, one can describe a system as large when it has a large mass, large size or, more generally, a large number of particles $N$. We are here taking a rather abstract approach, which will allow us to better understand the features that are crucial to distinguish macroscopic quantum states from other quantum states. Thus, from now on, we consider a state to be macroscopic in terms of the number of particles, which we assume to be large. In order to avoid the additional complexity of biological organisms like cats, we simplify the scenario by assuming that our large systems under study can be described as ensembles of quantum bits (qubits), which can generally represent quantum objects with one degree of freedom that can take two possible values, denoted as 0 and 1. Unlike classical bits, quantum bits can be in superposition states, like $\ket{+} \propto \ket{0} + \ket{1}$. To continue the analogy to Schr\"odinger's cat, we define now an oversimplified version of a cat that is just a system of $N \gg 1$ qubits, where each qubit represents one ``cell'' of the cat. The state $\ket{0}$ corresponds to the cell being dead and the state $\ket{1}$ to the cell being alive. This notion of a cat will accompany us throughout the entire manuscript and will be the object of study. We define different cats by means of different quantum states and we compare them in order to characterize which of them are truly macroscopic quantum state and which are not. To start with, we work with two examples that illustrate two extreme cases. This choice of extreme cases makes their comparison clearer, exposing the main features one should look at when studying the macroscopicity of quantum states.

Our first cat $A$ is represented by the GHZ state
\begin{equation}
     \label{eq:GHZ_state}
    \textrm{(A)}\,
    \ket{\rm{GHZ_N}} 
    = \frac{1}{\sqrt{2}} ( \ket{00..0} + \ket{11..1} ),
\end{equation} 
where $\ket{00..0}=\ket{0}^{\otimes N}$ and $\ket{11..1}=\ket{1}^{\otimes N}$. Cat $A$ is in a superposition of all its cells being dead ($\ket{0}^{\otimes N}$) and all its cells being alive ($\ket{1}^{\otimes N}$).

\begin{figure}
\centering
\includegraphics[width=3.1in]{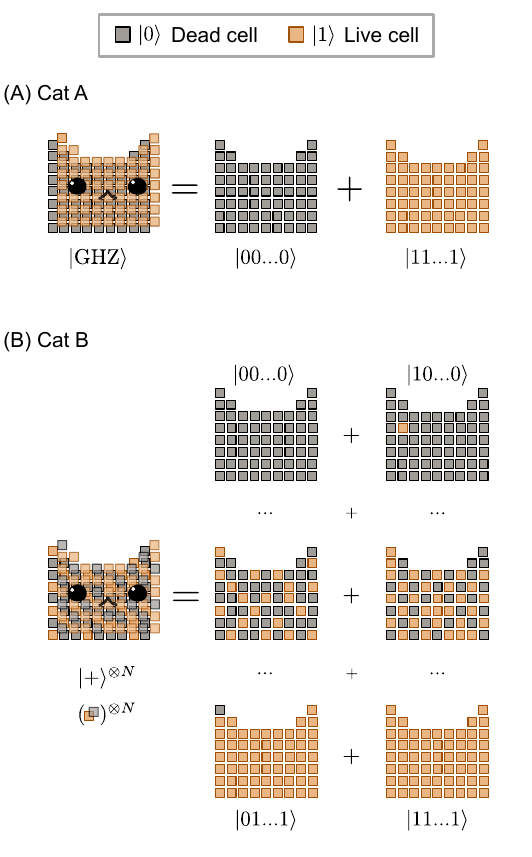}
\caption{Simplification of a cat as a system of $N$ qubits, where $N\gg 1$ is large. Each cell of the cat can be either dead ($\ket{0}$, grey square in the picture) or alive ($\ket{1}$, orange square), or in an arbitrary superposition of both states, as for instance $\ket{+} = \frac{1}{\sqrt{2}} (\ket{0}+\ket{1})$, which we depict by overlapping squares. Throughout this paper, we consider two different cats: (A) cat $A$, described as a GHZ state, i.e., as a superposition of all its cells being dead and all its cells being alive; and (B) cat $B$, described as $\ket{+}^{\otimes N}$, i.e., as a superposition of all the possible combinations of dead and live cells. See text for details.}\label{fig:catsAandB}
\end{figure}

Our second cat $B$ is represented by the following state
\begin{align} \label{eq:plusNstate}
    \textrm{(B)}\, \ket{+}^{\otimes N} &= \frac{1}{\sqrt{2}} (\ket{0} + \ket{1}) \otimes ...\otimes \frac{1}{\sqrt{2}} (\ket{0} + \ket{1}) \nonumber \\
     &= \frac{1}{\sqrt{2^N}} (\ket{00..0} + \ket{00..01} +  ...+\ket{11..1}),
\end{align}
which is a superposition of $2^N$ different states. Cat $B$ is in a superposition of all the possible combinations of live and dead cells, i.e. it contains the state with no live cells ($\ket{00..0}$), all the possible combinations with 1 live cell ($\ket{00..001}$, $\ket{00..010}$, ..., $\ket{10..000}$), all the possible combinations with 2 live cells, etc. Both cat $A$ and cat $B$ are depicted in Fig.~\ref{fig:catsAandB}. 

What stuck in everyone's mind from Schr\"odinger's thought experiment---and that is now even part of pop culture---is that his cat ended up being ``both dead and alive at the same time''. This sentence can also be said for both our cats, since both are in superpositions of dead and live cells. Then, in what sense are all these cats different? 

We will introduce a precise characterization of macroscopic quantum states that allows us to better describe states like our cats and to be able to point at what distinguishes them. At first sight, both cats can be naively described as macroscopically quantum, since both are macroscopic ($N$ is large) and both are made of quantum particles. In the following, we aim at explaining their differences in a pedagogical and visual way, in order to understand why we call only one of them a \textit{macroscopic quantum state}.

\begin{figure}
\begin{center}
\includegraphics[scale=0.6]{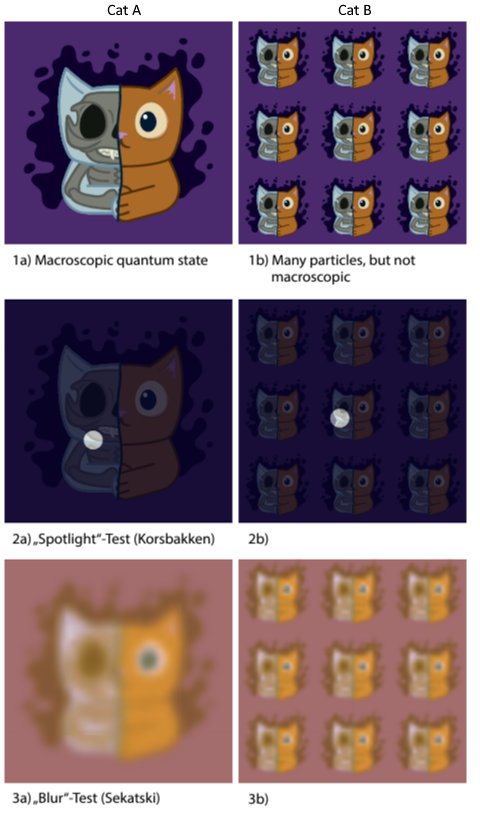}
\end{center}
\caption{1a) Visualization of the macroscopic quantum state (cat $A$) $|GHZ_N\rangle = \frac{1}{\sqrt{2}} ( |0 \rangle^{\otimes N} + |1 \rangle^{\otimes N})$. 1b) Visualization of non-macroscopic many-particle state (cat $B$) $|+\rangle^{\otimes N} = [\frac{1}{\sqrt{2}} ( |0 \rangle + |1 \rangle ) ]^{\otimes N}$. 2a) ``Spotlight"-test~\cite{korsbakken2007measurement} (Korsbakken at al.). Only for the macroscopic quantum state, a single-cell measurement is sufficient to characterize the whole state. 3a) ``Blur"-test~\cite{sekatski2014size} (Sekatski et. al.): How much blurring is allowed if one still wants to distinguish the states? For the macroscopic state, much more blurring is possible.} \label{fig:summary_intuition_measures}
\end{figure}

In broad terms, the main difference between cat $A$ and cat $B$ is that cat $A$ is in a superposition of two states that are strongly correlated at the collective level, whereas in cat $B$, the states of all cells are not correlated at all and one can describe the global superposition of many $N$-cell states also as a $1$-cell superposition that is then multiplied $N$ times, i.e. $\ket{+}^{\otimes N}$. Intuitively speaking, for a macroscopic state, the cat only has "one" life - a single measurement leads to a completely mixed state, see Fig \ref{fig:summary_intuition_measures}. For large, but not macroscopic quantum states, the cat has many lifes, as a single measurement destroys only s small subset of the superposition state. 

The authors in~\cite{frowis2012measures} start with a working definition that reads: `A quantum state is called macroscopic if it is capable of showing behaviour that is neither classical nor an accumulated microscopic quantum effect.' One can visualize this working definition in Fig.~\ref{fig:summary_intuition_measures} (top panels). Following this intuition, we can see cat $B$ as an accumulation of a microscopic quantum effect---and not as a true macroscopic quantum state, since the individual superposition $\ket{+}$ is ``accumulated'' $N$ times. This idea helps us introduce the notion of \textit{effective size} of the system, which is, intuitively, the scale at which quantum effects occur. Cat $B$ thus has an effective size of 1 cell, which we denote as $N_{\textrm{eff}}=1$, because it is the scale of the quantum system that is then being multiplied until a macroscopic scale is reached, whereas cat $A$ has an effective size of $N$, since it is not an accumulation of a microscopic quantum effect. In contrast to the product state of cat $B$, cat $A$ is a state where all N particles are entangled.

Determining the effective size of an arbitrary quantum state is not as easy as it seems. In our simple examples, we can directly see cat $B$ as an accumulation of microscopic quantum effects using the expression $\ket{+}^{\otimes N}$. However, one typically does not have access to such an expression. 

Therefore, several macroscopicity measures have been introduced over the past years to help determine whether a given quantum state is macroscopic or not. The intuition behind some of them is depicted in Fig.~\ref{fig:summary_intuition_measures} (lower panels). In the following section, we present the measures sketched in Fig.~\ref{fig:summary_intuition_measures} in more detail by applying them to our two cats $A$ and $B$.

% Give general properties of "macroscopic" quantum effect:

% For a macroscopic state: The Cat only has "one" life - a single measurement leads to a completely mixed state, see Fig \ref{fig:summary_intuition_measures}. For large, but not macroscopic quantum states like superconductors: The cat has many lifes, a single measurement destroys only a single cooper pair - Show that this is fulfilled for GHZ-state.

\section{Macroscopicity measures}\label{sec:macroscopicity_measures}

\subsection{Measures for superposition states $\propto \ket{\phi_1} + \ket{\phi_2}$}
In this section, we explain two different ways to determine whether a given quantum state is a macroscopic quantum state, or an accumulation of microscopic quantum states. We introduce two macroscopicity measures that can only be applied to superposition states of the form
\begin{equation}
    \ket{\psi} = \frac{1}{\sqrt{2}} (\ket{\phi_1} + \ket{\phi_2}).
\end{equation}

\textbf{Macroscopicity measure 1: Spotlight-test. }First, we start with the measure by Korsbakken et al.~\cite{korsbakken2007measurement}, which we also call Spotlight-test. Their measure is based on the intuitive idea that two macroscopic states, like a dead and a live cat, are easily distinguishable from each other. It is enough to measure one cell of the cat to determine whether the cat was alive or dead. However, this is not the case for quantum states that are large but not macroscopically quantum. 

This intuition is formalized in~\cite{korsbakken2007measurement} as follows. The effective size $N_{\textrm{eff}}$ of a quantum state $\ket{\psi}$ of $N \gg 1$ particles is given by
\begin{equation}
    N_{\textrm{eff}} (\ket{\psi}) = \frac{N}{n_\textrm{min}},
    \label{Nmin}
\end{equation}
where $n_\textrm{min}$ is the minimum size (in number of particles) of any subsystem of the initial state $\ket{\psi}$ that one needs to measure to collapse a superposition state $\propto \ket{\phi_1} + \ket{\phi_2}$ to one of its branches (either $\ket{\phi_1}$ or $\ket{\phi_2}$) with probability greater than a threshold value $p_d$. In order to obtain $n_\textrm{min}$, one first starts by measuring all the single-particle subsystems of the initial state, that is, $n=1$. If by performing these measurements, one cannot distinguish $\ket{\phi_1}$ from $\ket{\phi_2}$ with probability higher than $p_d$, then the process starts again with subsystems of two particles ($n=2$), then three particles, etc. 

We illustrate this measure by applying it first to our cat $A$, and then to our cat $B$. 

\begin{figure*}
\centering
\includegraphics[width=0.9\textwidth]{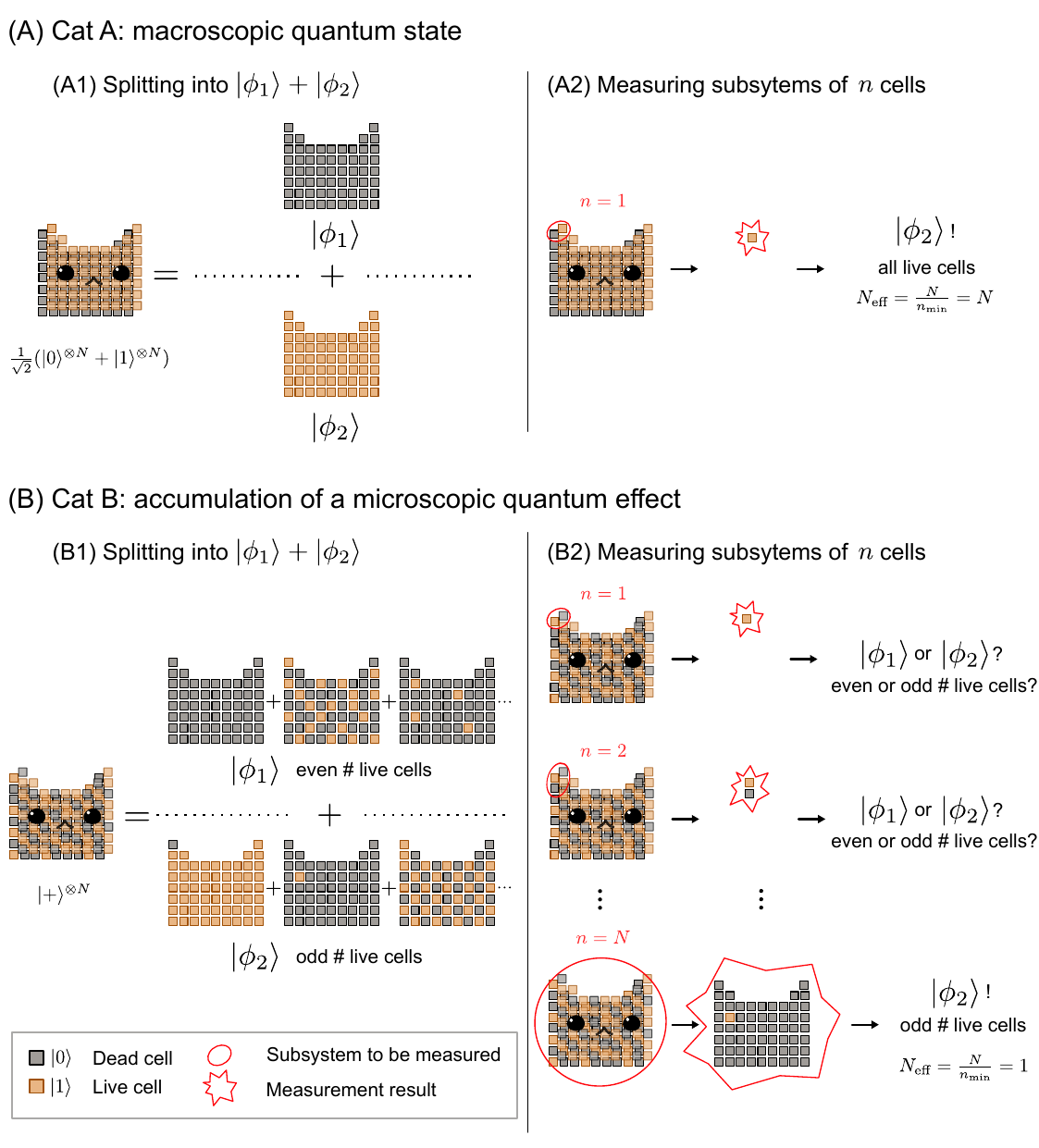}
\caption{We apply the Spotlight-test~\cite{korsbakken2007measurement} to cats $A$ and $B$, obtaining an effective size of $N_{\textrm{eff}} (\text{GHZ}) = \frac{N}{1} = N$ and $N_{\textrm{eff}} (\ket{+}^{\otimes N}) = \frac{N}{N} = 1$, respectively. Only cat $A$ is macroscopically quantum. (A1) The GHZ state already has two clear branches. (A2) Only by measuring one cell ($n_\textrm{min} = 1$) of the cat are we able to know whether the cat state has collapsed to either $\ket{00..0}$---all cells are dead---or $\ket{11..1}$---all cells are alive. (B1) First, we identify the two branches of the state, a first branch where all the states have an even number of live cells and a second branch where all the states have an odd number of live cells. (B2) Second, we perform measurements on subsystems of $n$ cells, starting with $n=1$ and increasing\footnote{Note that for each value of $n$, one needs a new cat $B$. That is, measurements are not done successively on the same cat.} $n$ until we are able to distinguish whether the state of the cat after the measurement has collapsed to the branch with even number of live cells or the branch with odd number of live cells. To do so, we need to measure the entire cat, that is, $n_\textrm{min} = N$.}\label{fig:korsbakken}
\end{figure*}

\textit{(A) GHZ state.---} Since this measure can only be applied to superposition states of the form $1/\sqrt{2} (\ket{\phi_1} + \ket{\phi_2})$, we begin by identifying the two branches. In this case, the GHZ state already has that form (Eq.~\eqref{eq:GHZ_state}) so it is enough to identify
\begin{align}
    \ket{\phi_1} = \ket{0}^{\otimes N}, \nonumber \\
    \ket{\phi_2} = \ket{1}^{\otimes N},
\end{align}
where we can directly see that one only needs to measure one particle to know if the state has collapsed to branch $\ket{0}^{\otimes N}$ (all cells are dead) or branch $\ket{1}^{\otimes N}$ (all cells are alive), i.e., $n_\textrm{min} = 1$. Thus, $N_{\textrm{eff}} (\text{GHZ}) = \frac{N}{1} = N$. The application of this measure to cat $A$ can be visualized in Fig.~\ref{fig:korsbakken}A, where we first identify the two branches and then start by measuring a single cell, which turns out to be sufficient to determine whether the cat has collapsed to the $\ket{\phi_1}$ or the $\ket{\phi_2}$ branch. If the cell was dead (alive), the cat has collapsed to the branch where all cells are dead (alive).

\textit{(B) $\ket{+}^{\otimes N}$ state.---} As we see in Eq.~\eqref{eq:plusNstate}, the state is not yet in a superposition of the form $1/\sqrt{2} (\ket{\phi_1} + \ket{\phi_2})$, so we first need to write it in such a form:
\begin{align}
    \ket{+}^{\otimes N} = \frac{1}{\sqrt{2}} &( \frac{1}{\sqrt{2}} 
 (\ket{+}^{\otimes N} + \ket{-}^{\otimes N}) \nonumber \\
 &+ \frac{1}{\sqrt{2}} (\ket{+}^{\otimes N} - \ket{-}^{\otimes N}) ),\label{eq:+_korsbakken}
\end{align}
from which we can identify the two branches,
\begin{align}
    \ket{\phi_1} = \frac{1}{\sqrt{2}} (\ket{+}^{\otimes N} + \ket{-}^{\otimes N}), \nonumber \\
    \ket{\phi_2} = \frac{1}{\sqrt{2}} (\ket{+}^{\otimes N} - \ket{-}^{\otimes N}).
\end{align}

By writing each branch of the state $\ket{+}^{\otimes N}$ in terms of states containing dead and live cells, one observes that branch $\ket{\phi_1}$ contains states with an even number of live cells and branch $\ket{\phi_2}$ states with an odd number of live cells. For example, for $N=3$, 
 \begin{align}
     \ket{\phi_1} \propto \ket{000} + \ket{110}+\ket{011}+\ket{101}, \\
     \ket{\phi_2} \propto \ket{111} + \ket{001}+\ket{010}+\ket{100}.
 \end{align}
This example is for a small number of particles but it can be easily computed and gives a good starting point to generalize the idea to $N \gg 1$. 

After noticing that the branches $\ket{\phi_1}$ and $\ket{\phi_2}$ correspond to states with even and odd number of live cells, respectively, we can directly observe that only by measuring all the cells of cat $B$ ($n_\textrm{min}=N$) can one collapse the cat state to either $\ket{\phi_1}$ or $\ket{\phi_2}$ (see Fig.~\ref{fig:korsbakken}B). Measuring e.g. just one cell does not give us any information about the total number of live cells, so we cannot know whether that number is even or odd. Therefore, $N_{\textrm{eff}} (\ket{+}^{\otimes N}) = \frac{N}{N} = 1$. 

After applying the Spotlight-test to both cat $A$ and cat $B$, we confirm the intuitions developed in Section~\ref{sec:effective_size}, i.e., cat $A$ is the only macroscopic quantum state. Even though the apparent size of cat $B$ is large (it contains a large number of cells $N$), the effective size where the quantum effects take place is just 1.

\begin{figure*}
\centering
\includegraphics[width=0.7\textwidth]{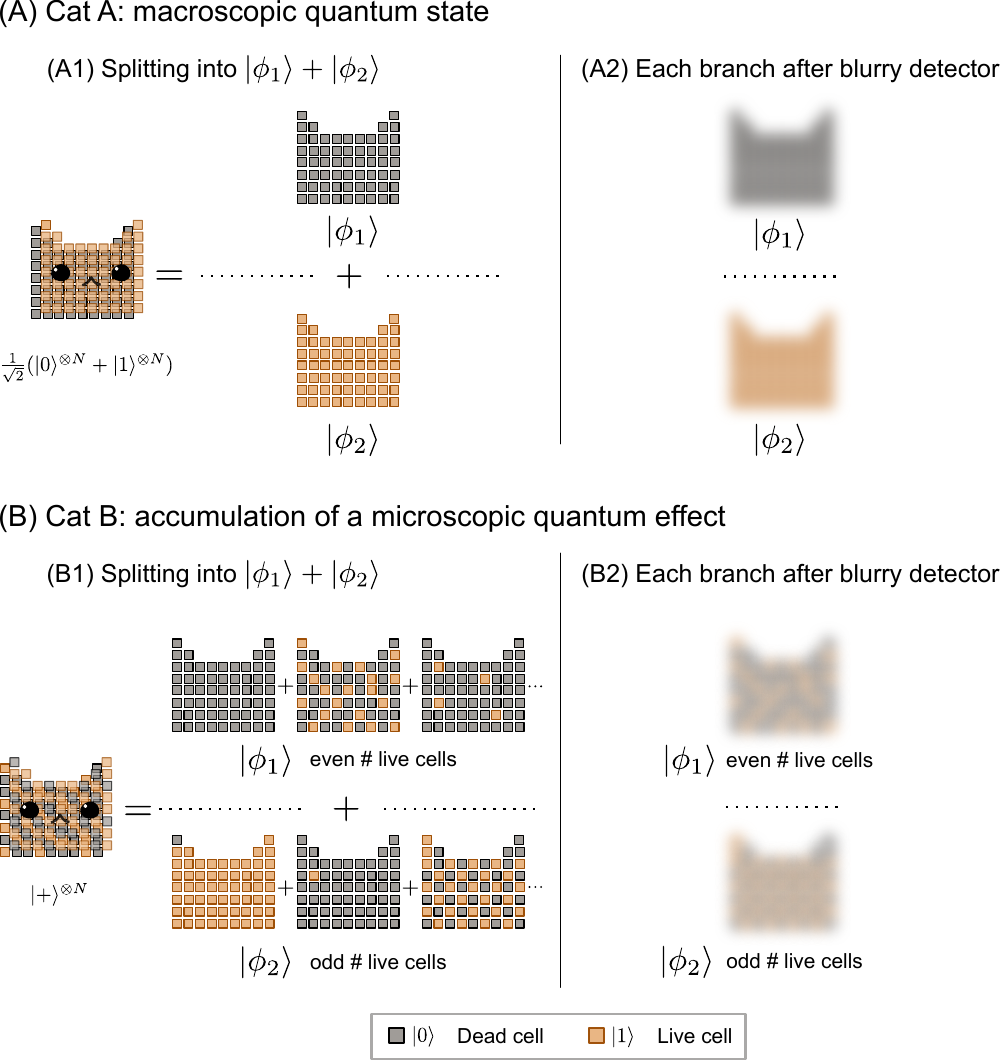}
\caption{We apply the Blur-test~\cite{sekatski2014size} to cats $A$ and $B$, and we see that the branches of the superposition are easily distinguishable after taking a blurry ``picture'' of cat $A$, but not after taking a blurry ``picture'' of cat $B$. With the same amount of noise (blurriness), we cannot say that the ``picture'' we got from cat $B$ corresponds to branch $\ket{\phi_1}$ or to branch $\ket{\phi_2}$. Thus, only cat $A$ is macroscopically quantum. (A1) The GHZ state already has two clear branches. (A2) Result of how each branch looks like after taking a blurry ``picture'' and obtaining that branch. (B1) First, we identify the two branches of the state, a first branch where all the states have an even number of live cells and a second branch where all the states have an odd number of live cells. (B2) Result of how each branch looks like after taking a blurry ``picture'' and obtaining that branch. Note that, for branch $\ket{\phi_1}$, all the possible configurations with an even number of live cells overlap. Same for branch $\ket{\phi_2}$.}\label{fig:sekatski}
\end{figure*}

\textbf{Macroscopicity measure 2: Blur-test.} This measure~\cite{sekatski2014size} was developed for photonic systems. However, it can also be applied to ensembles of qubits. Here, we only use the intuitive idea of the measure and apply it to our two cats $A$ and $B$. In a similar spirit to the Spotlight-test, the idea of the Blur-test is that two macroscopic states, like a dead and a live cat, can be distinguished even if we measure them with a noisy classical detector. We represent the detector as a camera that takes a ``picture'' of our quantum cats, and the noise as the camera glass being blurry. True macroscopic states tolerate a large amount of blurriness and they can still be discriminated. In Fig.~\ref{fig:sekatski}, we use a really blurry camera glass and see that only the pictures of the branches of cat $A$ can be distinguished from each other., but not the branches of cat B. Indeed, if we are given a picture of cat $B$ with such blurriness, we cannot say if it corresponds to branch $\ket{\phi_1}$ (even number of live cells) or to branch $\ket{\phi_2}$ (odd number of live cells). To be able to count the number of live cells and say whether it is even or odd, one would need a camera with much better resolution.

\subsection{Quantum Fisher Information}\label{sec:QFI}
Up until now, we have given a basic intuition of the notion of effective size in Section~\ref{sec:effective_size} and we have presented a pedagogical way of explaining measures of macroscopicity that can be applied to superposition states of the form $\ket{\phi_1} + \ket{\phi_2}$ (Section~\ref{sec:macroscopicity_measures}). Based on the intuition gained in the previous sections, we here introduce a general macroscopicity measure---it can be applied to any quantum state---that relies on the quantum version of a statistical measure called Fisher Information.

The Quantum Fisher Information (QFI) of a pure quantum state $\ket{\psi}$ with respect to an observable $D=\sum_{i=1}^N D^{(i)}$---where each term $D^{(i)}$ is the local observable that acts on each particle $i$---is given by
\begin{equation}
    \mathcal{F}(\ket{\psi}, D) = 4 (\Delta_{\ket{\psi}} D)^2 = 4 (\langle D^2 \rangle_\psi - \langle D \rangle_\psi^2), \label{eq:QFI_purestates}
\end{equation}
where $(\Delta_{\ket{\psi}} D)^2$ is the variance of observable $D$ with respect to state $\ket{\psi}$ and $\langle D \rangle_\psi = \langle \psi |D| \psi \rangle$ is the expected value of observable $D$ when the state $\ket{\psi}$ is measured multiple times (see Appendix~\ref{app:QFI_pure_states} for a derivation of equation (\ref{eq:QFI_purestates}) from the general definition of QFI for general mixed states). For some observable $D$, macroscopic quantum states like the GHZ-state have larger variance, and in turn, a larger QFI, than quantum states that are not truly macroscopic. 

The observation that certain quantum states like the GHZ state have a large variance has been exploited in the context of metrology, where these states have been used as a resource to estimate a parameter with higher precision. We elaborate more on this in Section~\ref{sec:macrostates_and_metrology}. The capacity of macroscopic quantum systems of improving the precision in a metrology setup can be used as a criterion for macroscopicity. Fr\"owis et al.~\cite{frowis2012measures} linked the field of metrology to macroscopicity by defining a macroscopicity measure that is based on the QFI. According to this measure, a quantum state $\ket{\psi}$ is macroscopic if there exists an observable $D = \sum_{i=1}^N D^{(i)}$--- where each term $D^{(i)}$ is the local observable that acts on particle $i$---such that the QFI $\mathcal{F}(\ket{\psi}, D) = O(N^2)$. The effective size  for pure states is then given by
\begin{equation}
    N_\textrm{eff}(\ket{\psi}) = \max_{D:\textrm{local}} \frac{\mathcal{F}(\ket{\psi}, D)}{4N} =\max_{D:\textrm{local}} \frac{(\Delta D_{\ket{\psi}})^2}{N}  ,
    \label{Neff}
\end{equation}
where the QFI is normalized in such a way that $1\leq N_\textrm{eff} (\ket{\psi}) \leq N$. This expression generalizes the definition in Eq.~\eqref{Nmin} to any given quantum state. Truely macroscopic quantum states reach $N_\textrm{eff} (\ket{\psi})=O(N)$. 

We again resort to our two examples to illustrate this idea: (A) the macroscopic quantum state $\ket{\textrm{GHZ}}$ and (B) the accumulation of microscopic quantum states $\ket{+}^{\otimes N}$. We choose the observable $D$ to be $D=\sum_{i=1}^N \sigma_z^{(i)}$. 

\begin{figure*}[!]
\centering
\includegraphics[width=\textwidth]{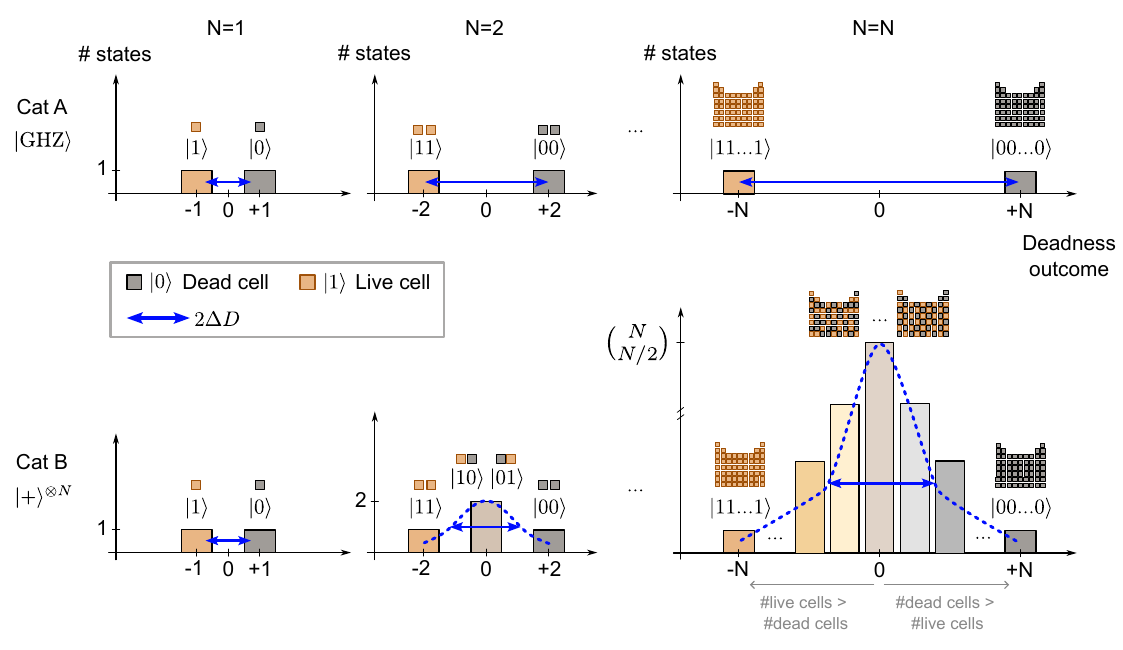}
\caption{Visualization of how the variance $(\Delta D)^2$ of the cat ``deadness'' ($D=\sum_{i=1}^N \sigma_z^{(i)}$) scales with the number of cells $N$. True macroscopic quantum states like the GHZ state show a quadratic scaling in the variance---thus, also in the QFI---with $N$, since there are only two possible outcomes: either the cat is fully alive (outcome $-N$) or fully dead (outcome $+N$), which are the extremes of the ``deadness'' spectrum. In contrast, non-macroscopic quantum states like $\ket{+}^{\otimes N}$ show a linear scaling with $N$ because all the possibilities of dead and live cells are present in the superposition, which leads to outcomes that are more concentrated around the average value $\mu=0$. Note that this is a sketch for visualization, the scales of the plot do not correspond to the actual scales.}\label{fig:visual_QFI_variance}
\end{figure*}

Macroscopic quantum states have a variance (and a QFI) that scales quadratically with the number of particles $N$. The difference in the QFI scaling between our two example states can be visualized in Fig.~\ref{fig:visual_QFI_variance}, where we can visualize how, as the number of cells increases, the variance of the cat ``deadness'' increases as $N^2$ for the GHZ state, whereas it only increases as $N$ for state $\ket{+}^{\otimes N}$. In the latter case, there are not quantum correlations between particles (no entanglement), so the possible outcomes are more concentrated around the average value $\mu =0$, since all the possible combinations of dead and live cells are present in the superposition $\ket{+}^{\otimes N}$ (see Eq.~\eqref{eq:plusNstate}). In the GHZ state, the state of all particles is strongly correlated (particles are entangled) so there are only two possible outcomes: either the cat is fully alive (outcome $-N$) or fully dead (outcome $+N$), which are the extremes of the ``deadness'' spectrum. In what follows, we will elaborate this intuitive idea more in detail.

Following the framework introduced in Section~\ref{sec:effective_size}, we interpret measuring observable $D$ as the process of measuring the ``deadness'' of the cat. The local observable $\sigma_z^{(i)} = \ketbra{0}{0} - \ketbra{1}{1}$ that acts on each qubit $i$ as
\begin{align}
    \sigma_z \ket{0} = (+1) \ket{0}, \nonumber \\
    \sigma_z \ket{1} = (-1) \ket{1},
\end{align}
represents measuring the ``deadness'' of each cell, with outcomes $+1$ if the cell is dead and $-1$ if it is alive. The overall ``deadness'' $D$ of the cat is thus the sum over all cells ``deadness'', which has a maximum value $+N$ if all cells are dead and a minimum value $-N$ if all cells are alive. A ``deadness'' of $0$ means that there is the same number of live and dead cells.

\textit{(A) GHZ state.---} When we measure the ``deadness'' of the cat, we can get two possible outcomes, $+N$ or $-N$, with equal probability. There is only one state ($\ket{00..0}$) that can give outcome $+N$ in the superposition $\ket{\textrm{GHZ}}$:
\begin{align}
    \sum_{i=1}^N \sigma_z^{(i)} \ket{0}^{\otimes N} &= \sigma_z^{(1)}\otimes \id \otimes ...\otimes \id \ket{00..0} \nonumber \\
    &+ \id \otimes \sigma_z^{(2)}\otimes \id \otimes ...\otimes \id \ket{00..0} + ... \nonumber \\
    &= (+1)\ket{00..0} + ... + (+1) \ket{00..0} \nonumber \\
    &= (+N) \ket{00..0}.
\end{align}
Analogously, state $\ket{11..1}$ gives outcome $-N$. The variance of the cat ``deadness'', given by $(\Delta D)^2 = 1/s \sum_{i=1}^s (x_i - \mu)^2$, measures how spread the outcomes $x_i$ are with respect to the average value, which in this case is $\mu = (N-N)/2=0$. The value $s$ is the total number of states that form the superposition of the quantum cat. For the GHZ state, $s=2$ (all cells are alive or all are dead). Therefore, the variance of the cat ``deadness'' when we measure cat $A$ is
\begin{equation}
    (\Delta_{\ket{\textrm{GHZ}}} D)^2 = \frac{1}{2} ( (N-0)^2 + (-N-0)^2) = N^2.
\end{equation}
One obtains the same result with Eq.~\eqref{eq:QFI_purestates} (see Appendix~\ref{app:variances}).

\textit{(B) $\ket{+}^{\otimes N}$ state.---} As we see in Eq.~\eqref{eq:plusNstate}, state $\ket{+}^{\otimes N}$ contains all the possible combinations of dead and live cells. Unlike the GHZ state, in this state there are many states that give the same outcome when we measure the ``deadness'' of the cat. Table~\ref{table:plusN_statesandoutcomes} shows all the possible outcomes with their corresponding states. There are many states that have a low ``deadness'' value, thus the outcome values are more concentrated around the average value $\mu=0$, which in turn leads to a smaller variance than in the case with a GHZ state. Note that the GHZ state has only two states and they are far apart at the extremes of the ``deadness'' spectrum. 

\begin{table*}[htb!]
\begin{tabular}{c|c|c|c}
\# Live cells & Outcome & States                               & \# States        \\ \hline
0             & $+N$    & $\ket{00..0}$                        & 1                \\
1     & $(+1)\cdot (N-1) + (-1)\cdot 1 = N-2$ & $\ket{00..001}$, $\ket{00..010}$, ... , $\ket{10..000}$ & $\binom{N}{1} = \frac{N!}{1!\cdot (N-1)!}=N$   \\
...           & ...     & ...                                  & ...              \\
$N/2$         & 0       & $\ket{00..11}$, ... , $\ket{11..00}$ & $\binom{N}{N/2}$ \\
...           & ...     & ...                                  & ...              \\
$N-1$ & $(+1)\cdot 1 + (-1)\cdot (N-1) = -N+2$ & $\ket{11..110}$, $\ket{11..101}$, ... , $\ket{01..111}$ & $\binom{N}{N-1} = \frac{N!}{(N-1)!\cdot 1!}=N$ \\
$N$           & $-N$    & $\ket{11..1}$                        & 1               
\end{tabular}
\caption{Possible outcomes, and their corresponding states, when measuring the cat ``deadness'' ($D=\sum_{i=1}^N \sigma_z^{(i)}$) of state $\ket{+}^{\otimes N}$.}
\label{table:plusN_statesandoutcomes}
\end{table*}

The variance of the cat ``deadness'' for the $\ket{+}^{\otimes N}$ state is
\begin{align}
    (\Delta_{\ket{+}^{\otimes N}} D)^2 &= \frac{1}{2^N} (1\cdot N^2 + \binom{N}{1}\cdot (N-2)^2 \nonumber \\ 
    &+ ... + \binom{N}{N/2}\cdot (0)^2 + ... \nonumber \\
    &+\binom{N}{N-1}\cdot (-N+2)^2 +1\cdot (-N)^2 ) = N,
\end{align}
where the total number of states is $2^N$ and each summand has the form `\#States with a given Outcome' $\cdot$ `Outcome'$^2$. The final result of $N$ is not straightforward, but can be easily obtained using Eq.~\eqref{eq:QFI_purestates} (see Appendix~\ref{app:variances}).

Following this definition and Eq.~\eqref{eq:QFI_purestates}, our two examples have effective sizes
\begin{align}
    \textrm{(a)} \, N_\textrm{eff}(\ket{\textrm{GHZ}}) & =\max_{D:\textrm{local}} \frac{(\Delta D_{\ket{\psi}})^2}{N}  = \frac{4N^2}{4N} =N, \nonumber \\
    \textrm{(b)} \, N_\textrm{eff}(\ket{+}^{\otimes N}) & =\max_{D:\textrm{local}} \frac{(\Delta D_{\ket{\psi}})^2}{N}  = \frac{4N}{4N} =1.
\end{align}

The observation that certain quantum states like the GHZ state have a large variance has been exploited in the context of metrology, where these states have been used as a resource to estimate a parameter with higher precision. Indeed, the capacity of macroscopic quantum systems of improving the precision in a metrology setup can be used as a criterion for macroscopicity.  In this spirit, Fr\"owis et al.~\cite{frowis2012measures} linked the field of metrology to macroscopicity by defining a macroscopicity measure that is based on the QFI. 
We elaborate more on this in Section~\ref{sec:macrostates_and_metrology}.

\section{Hands-on activity: ``Create your macroscopic quantum cat''}\label{sec:hands-on-activity}

\begin{figure}
\begin{center}
\includegraphics[width=3.3in]{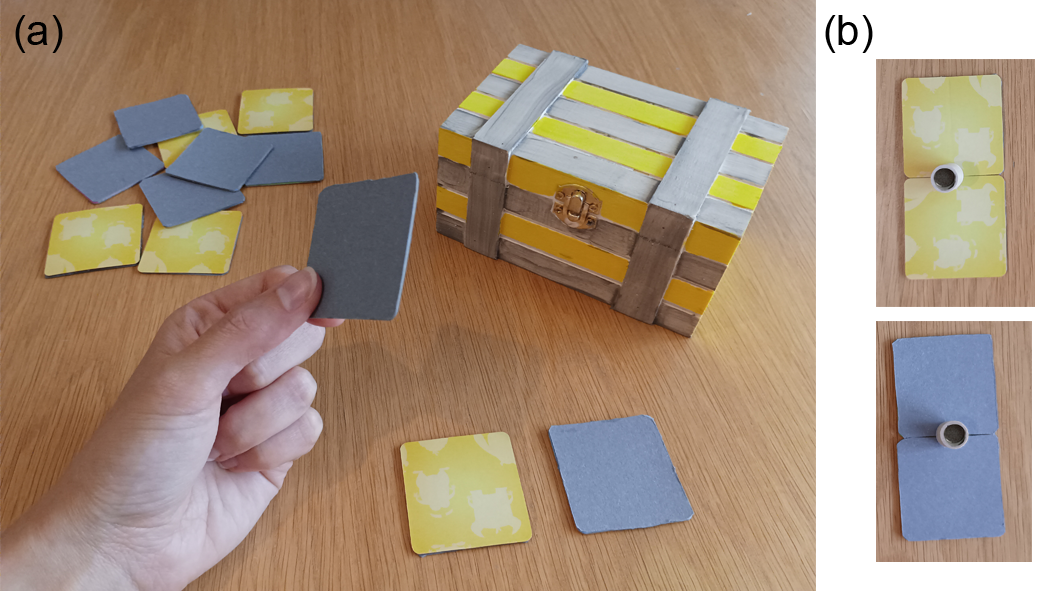}
\end{center}
\caption{(a) Box containing the cells to build our quantum cats. Each cell is in a superposition of being alive (yellow side) and dead (grey side), which is visually emphasized when we hold it in the air. (b) Two or more cells can be entangled by joining them along the sides with magnets.} \label{fig:box_quantum_cats}
\end{figure}

In this section, we develop a hands-on activity based on the visualization of cats with live and dead cells presented in the previous sections. In the box (see Fig.~\ref{fig:box_quantum_cats}), the student can find cells that are in a superposition of being alive and dead and the goal is to create a given type of quantum cat with those cells by combining and entangling them in different ways.

Each cat cell is represented by a squared piece of cardboard with two sides, one yellow (alive) and one grey (dead) and is considered to be in a superposition of the two sides, i.e. of being dead and alive, which becomes visually clear when we hold the cardboard in the air. Two or more cells can be entangled by joining them with the circular magnets along their sides (see Fig.~\ref{fig:box_quantum_cats}b). Note that once two or more cells are entangled with the magnets, they will keep together if we pick them up and hold them in the air, showing the superposition of all of them being alive and all of them being dead. 

\begin{figure}
\begin{center}
\includegraphics[width=3.2in]{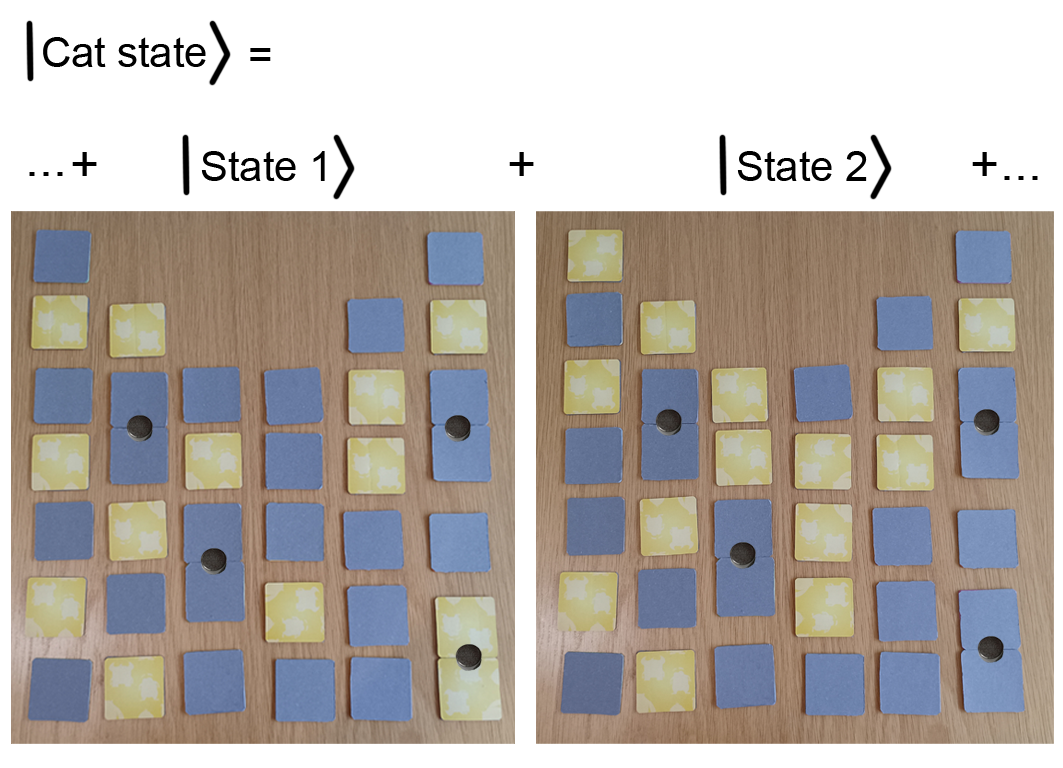}
\end{center}
\caption{Example of a cat and some of the states that are in its global superposition state.} \label{fig:configurations_quantum_cat_box}
\end{figure}

The creation of a cat from these cells goes as follows: 
\begin{enumerate}
    \item First, you can create the shape of the cat by placing the cells on the table. The side that is facing upwards does not matter, and the reason for this will become clear in step 3.

    \item Second, you can decide which cells to entangle by joining them with the magnets.

    \item Third, you can see all the possible states that belong to the superposition state of the cat you created by flipping around the cells (see Fig.~\ref{fig:configurations_quantum_cat_box}). Note that cells that are entangled flip together, so the more cells you entangle, the less (different) states your cat has in the superposition.

    \item Fourth, you can measure the state of your cat by taking a picture of the table with the cells on it at a given instant. Even though the cat was in the superposition of all the possible flipping combinations (which you did yourself by flipping the cardboard pieces), you can only capture one configuration with your camera.
\end{enumerate}

\begin{figure}
\begin{center}
\includegraphics[width=3.2in]{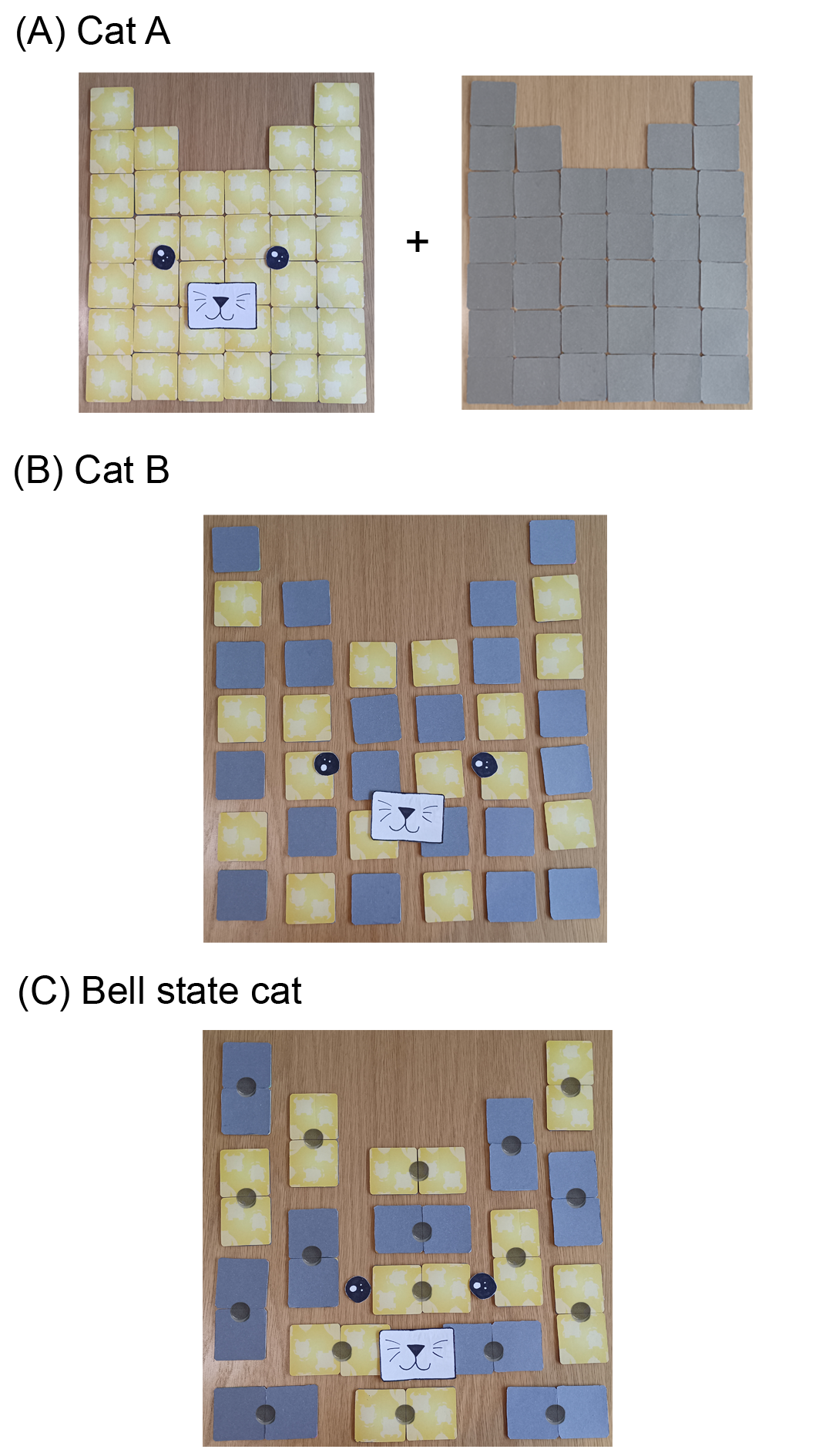}
\end{center}
\caption{Cell configurations that correspond to some of the cats we present in this paper. (A) Cat A - $\ket{\textrm{GHZ}}$ (magnets have been omitted in the image for clarity, but all cells are entangled), (B) Cat B - $\ket{+}^{\otimes N}$ - in one of its superposition states, (C) Bell state cat - $\ket{\Phi^+}^{\otimes N/2}$ - in one of its superposition states.} \label{fig:our_quantum_cats_box}
\end{figure}

In Fig.~\ref{fig:our_quantum_cats_box}, we show how the configurations of the cats presented in previous sections would look like. In the configuration of Cat A (Fig.~\ref{fig:our_quantum_cats_box}A), which is the macroscopic quantum state $\ket{\textrm{GHZ}_N}$, all cells are entangled, so we could hold the entire cat up in the air and see the superposition of all cells alive ($\ket{11..1}$) on one side and all cells dead ($\ket{00..0}$) on the other side.

\begin{figure}
\begin{center}
\includegraphics[width=3.4in]{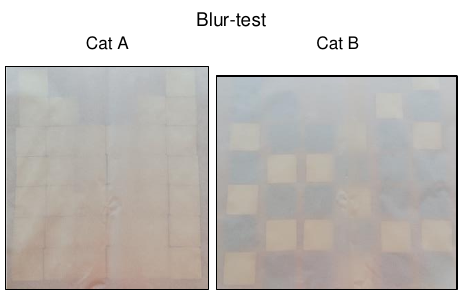}
\end{center}
\caption{The blur-test can be performed by placing a translucent paper in between the cardboard cells of the cat and our eyes. We see how we can easily see that Cat A is in the branch ``all cells alive'', whereas it is not clear in which branch (odd or even total number of live cells) Cat B is, because it is hard to distinguish every single cell, and thus, it is hard to count the total number of live cells.} \label{fig:blur_test_activity}
\end{figure}

We suggest using this activity to introduce the main ideas in macroscopicity and the macroscopicity measures explained in Sec.~\ref{sec:macroscopicity_measures}. For example, the Spotlight-test can be introduced by first creating our cat A and, before giving any information about the actual measure, asking how many cells the student should measure (take a picture) to know which of the two states of the superposition they got. Cat B can also be created and analyzed, counting e.g. how many different states it has in its superposition ($2^N$), and `splitting' the superposition into two groups, one with all the states that have an odd number of live cells and another one with all the states that have an even number of live cells. Then, analogously to cat A, the Spotlight-test can be introduced by asking how many cells the student should measure to determine whether the state they got has an odd or an even number of cells.

The Blur-test can also be performed by looking at the created cat through a translucent paper (see Fig.~\ref{fig:blur_test_activity}). Similarly to Fig.~\ref{fig:sekatski}, we see that the branches of the superposition are only easily distinguished in cases such as Cat A (macroscopic).

This activity provides a way to introduce the topics in a way in which the student has a main active role and the teacher can lead a discussion assisted by the activity material that enhances the student's critical thinking. After students get familiar with the material and create some cats, the discussion on the topic of macroscopicity can start. Some examples of guiding questions are:
\begin{itemize}
    \item What is the difference between the cats you created?
    \item All of them have the same number of cells, so how would you characterize them based on the differences you observed? Are they similar or different to Schr\"odinger's cat?
    \item When you measure your cat, how many possible outcomes do you get? And when you measure cat~A? 
    \item If you take many pictures of your cat, in the different moments after you flip some cells, and for each picture you compute $D = (+1)\times \textrm{\# of dead cells} + (-1)\times \textrm{\# of live cells} $. How is the variance of $D$ when you measure your cat compared to the variance of $D$ when you measure cat A? 
\end{itemize}

\section{Transition between macroscopic and non-macroscopic quantum systems}\label{sec:transition}

In this section, we discuss several less obvious examples, where a smooth transition between microscopic and macroscopic behaviour depending on some parameter (like the squeezing parameter) can be observed. For each of these examples, we also discuss possible applications. 

\subsection{Transition from Cat B to Cat A}

We start with the Cat state B and want to find a transition to Cat state A. Cat B is a product of $N$ single-qubit states $\ket{+} =\frac{1}{\sqrt{2}} ( \ket{0} + \ket{1} ) $. We view the latter single-qubit state as a ``GHZ'' state, defined as  $\ket{\textrm{GHZ}_p} = \frac{1}{\sqrt{p}} ( \ket{00..0}_p + \ket{11..1}_p )$ with $p=1$, where $\ket{00..0}_p = \ket{0}^{\otimes p}$ and $\ket{11..1}_p = \ket{1}^{\otimes p}$.  For $p=2$, we obtain a Bell-state, and for $p>2$, GHZ states in the original meaning of multi-particle entanglement emerge.  

\begin{figure}
\begin{center}
\includegraphics[width=3.3in]{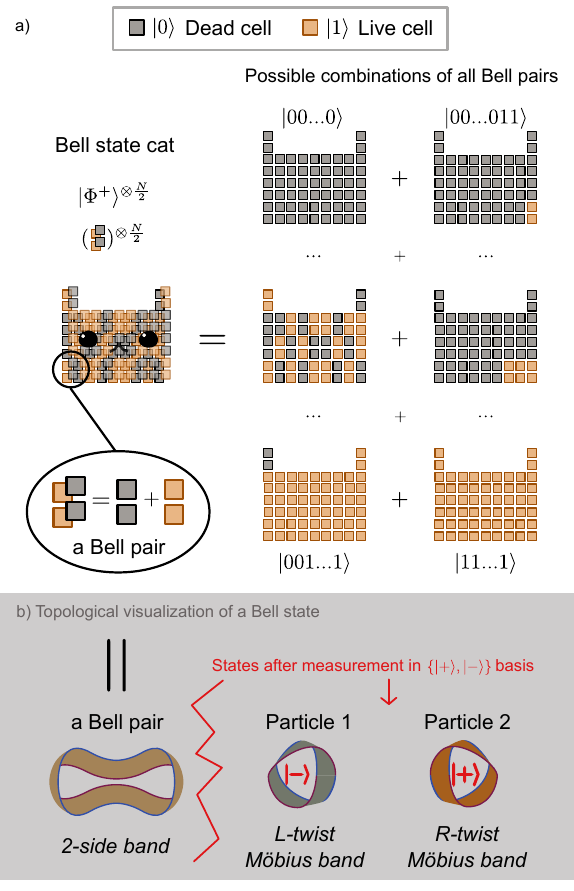}
\end{center}
\caption{a) Depiction of a Bell-state cat $|\Phi^+ \rangle^{\otimes N/2}$, consisting of $N\gg 1$ cells grouped in $N/2$ pairs of particles that are in a Bell state $\ket{\Phi^+} = 1/\sqrt{2} (\ket{00}+\ket{11})= 1/\sqrt{2} (\ket{+- }+\ket{- +})$. b) Each Bell pair can be topologically visualized as a single 2-side band, which further emphasizes that the two entangled particles of the pair cannot be described independently from each other. Upon measurement in the $\{\ket{+}, \ket{-} \}$ basis, the 2-side band splits into two individual M\"obius strips, i.e. particles are no longer entangled (see \cite{topology} for details). Only one ($\ket{-+}$) of the two possible measurement outcomes is depicted.} \label{fig:Cooper}
\end{figure}

We consider (for the case that $N/p$ is integer) the class of product states with $N/p$ identical copies of the state $ \ket{\textrm{GHZ}_p} $, that is,
\bea
|\textrm{GHZ}_p\rangle^{\otimes N/p} = \frac{1}{p^{N/(2p)}} ( \ket{00..0}_p + \ket{11..1}_p)^{\otimes N/p}. \label{eq:general_GHZ_p}
\eea
In such a way, we find a transition from Cat state B to Cat state A, with $N_{\rm eff}=p$. As an example of particular physical importance, we consider the product of $N/2$ Bell-states ($p=2$), that is
\bea \label{eq:Bell_state}
|\Phi^+ \rangle^{\otimes N/2} = \frac{1}{\sqrt{2^{N/2}}} ( |0 0\rangle + |1 1 \rangle)^{\otimes N/2}  \\ \nonumber
= \frac{1}{\sqrt{2^{N/2}}} ( |+ - \rangle + |- + \rangle)^{\otimes N/2},
\eea
which is $not$ a macroscopic quantum system, but an accumulated microscopic quantum system with $N_{\textrm{eff}}=2$. Note that the same state $|\Phi^+\rangle$ has two slightly different expressions depending on whether we write it in the $\{\ket{0},\ket{1}\}$ basis or in the $\{\ket{+},\ket{-}\}$ basis. We depict the Bell-state cat in Fig. \ref{fig:our_quantum_cats_box}C and in Fig.~\ref{fig:Cooper}. Note that, in contrast to cat B (Fig.~\ref{fig:catsAandB}B), in the Bell-state cat cells are grouped in pairs of entangled particles, which we emphasize with the topological visualization of a Bell state as a single, 2-side closed band~\cite{topology}. This band represents two entangled particles whose states are correlated in such a way that they form this single topological object with no ``twists''. This topology illustrates the fact that entangled particles cannot be described individually, i.e. independently of one another. However, upon measurement, particles become independent and are no longer entangled. In the topological analogy, this is represented by the single 2-side band breaking into two independent M\"obius strips, one with an R-twist and another one with an L-twist (one for each particle, see Fig.~\ref{fig:Cooper}). The R-twist M\"obius strip represents state $\ket{+}$, and the L-twist state $|-\rangle = 1/\sqrt{2} (\ket{0}-\ket{1})$. This picture thus symbolizes a measurement in the $\{\ket{+},\ket{-}\}$ basis, upon which the two particles can be found either in the $\ket{+-}$ state or in the $\ket{-+}$ state (see Eq.~\eqref{eq:Bell_state}, and note that the measurement results are correlated in both bases $\{\ket{0},\ket{1}\}$ and $\{\ket{+},\ket{-}\}$). Overall, with the topological visualization, the cat consists of $N/2$ closed, 2-side bands.

\begin{figure}
\begin{center}
\includegraphics[width=2.2in]{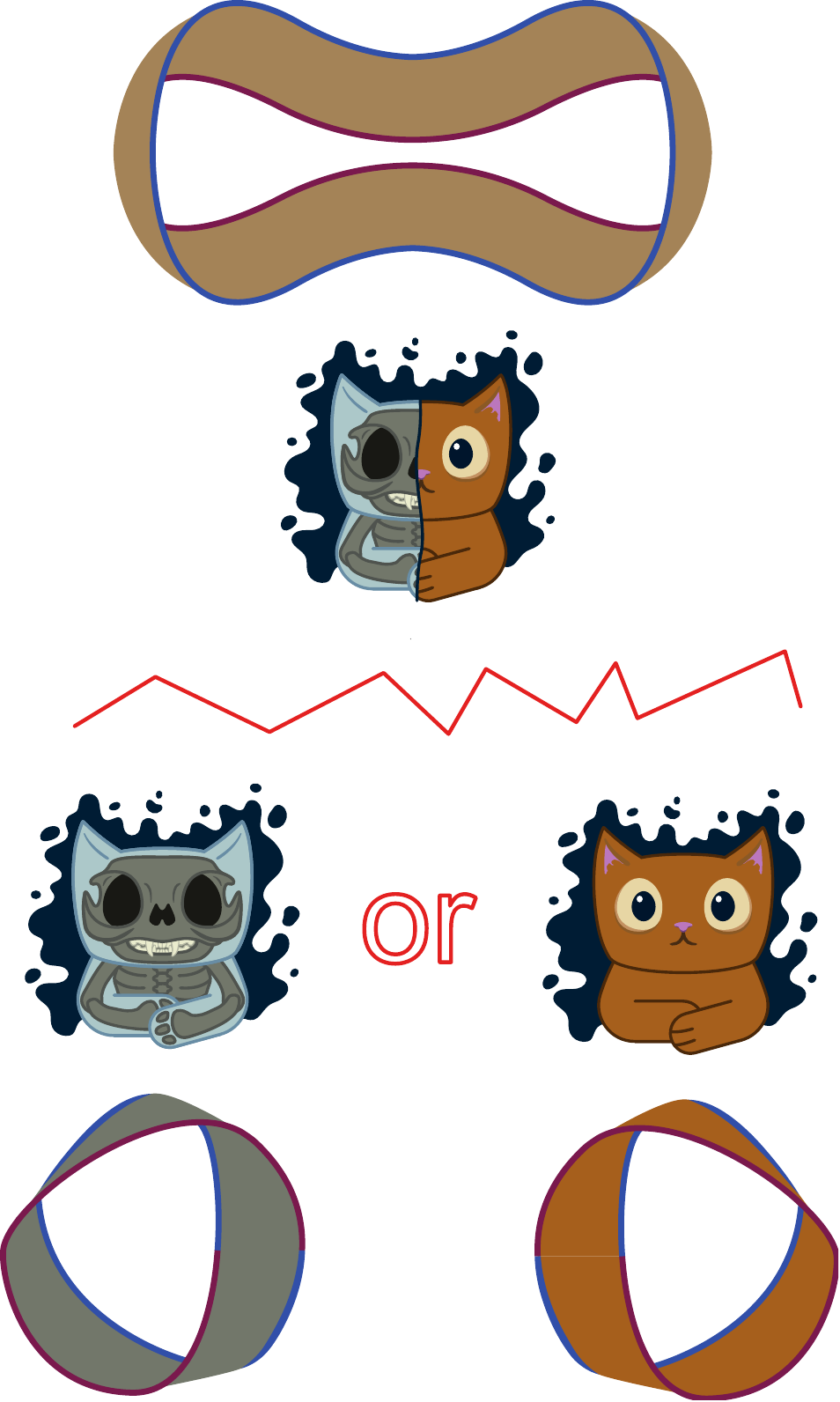}
\end{center}
\caption{We recover cat A by setting $p=N$ in Eq.~\eqref{eq:general_GHZ_p}. Its topological visualization is one big 2-side band that represents the entire cat. Upon measurement, the cat can be found in one of the two branches of the superposition, represented by one big L-twist (dead cat) and one big R-twist (alive cat) M\"obius strips.} \label{fig:extreme_p=N_macro}
\end{figure}

By setting $p=N$, we obtain cat A again, which is the extreme case in which there is one large group of $N$ particles, all in superposition (see Fig.~\ref{fig:catsAandB}A). In Fig.~\ref{fig:extreme_p=N_macro}, we see the topological visualization of cat A as one big 2-side band that, when measured, splits into a big L-twist and a big R-twist M\"obius strips, and we can only get one of the two as a result of the measurement. 

\textit{Applications:} For $p=2$ (accumulation of Bell pairs), we may think of Cooper pairs in superconductors. In such a case, we see that Cooper pairs in superconductivity lead to accumulated microscopic quantum effects, even if the wave function itself is macroscopic. For example, consider e.g. Andreev reflection at the boundary between a normal and a superconductor. In this process, a single electron from the normal conductor scatters with another electron directly at the interface of the  superconductor, creating an entangled Cooper pair in the superconductor. The ``missing electron'' at the interface can be viewed as a ``hole'' that is backscattered in the normal conductor. The opposite process is also possible: if a hole in the normal conductor hits the interface of the superconductor, a Cooper pair in the superconductor breaks into two electrons, one filling the hole, the other being backscattered into the normal conductor following the path of the original hole with opposite momentum. This microscopic quantum effect only affects a single Cooper pair, which is independent of the bulk. In our picture from Fig. \ref{fig:Cooper}, we can consider each Bell pair as a Cooper pair and the entire cat as the bulk of the superconductor. We thus may view the single backscattering event as a breakdown of a single Cooper pair, leaving all the other Cooper pairs unaffected. In a seminal paper, Leggett argued that superconductivity itself is a paradigmatic example for a microscopic quantum effect \cite{Leggett1980}, however, he also raised the question of how a superposition of left- and right moving currents in a SQUID should be interpreted. As reviewed in \cite{Frowis2017}, this question is still under debate, even 40 years after it was raised. One reason may be that several different theoretical approaches can be used to formalize the problem. One can show the equivalence of the system to a massive particle moving in a tilted double-well potential, when considering collective degrees of freedom such as (i) the total flux in the SQUID, or (ii) the total left- and right moving current. If one considers the latter option, and interprets the superposition of the left- and the right moving superconducting currents as a macroscopic effect, the corresponding picture is that of Fig.~\ref{fig:extreme_p=N_macro}. When the current is measured, eventually, the superposition breaks down and either the left (L-twist M\"obius band) or the right (R-twist M\"obius band) current is observed. However, on the microscopic scale, there are many Cooper pairs involved, as shown in Fig. \ref{fig:Cooper}. If rather than measuring the total current, only single Cooper pairs are affected, as in e.g. Andreev reflection, only microscopic effects will emerge. Thus, depending on the perspective chosen and the chosen degrees of freedom to be measured, different answers will be obtained to the question whether the system is macroscopic or not. This example illustrates the complexity of the problem, and makes evident that further research is necessary to resolve the problem of macroscopicity.

\subsection{Generalized GHZ states (Cat A)}

In this section, rather than starting with the microscopic Cat state B to find a transition to the macroscopic Cat state A, we do the opposite and start with macroscopic Cat state A and detune it to a microscopic state. We introduce the single-qubit superposition state
\begin{equation}
\ket{\epsilon}= \cos(\epsilon)  \ket{0} + \sin(\epsilon)  \ket{1}.
\end{equation} 
Note that for $\epsilon = \pi/4$, we recover the $\ket{+}$ state. We define the generalized GHZ state as
\begin{equation}
    \ket{\rm{GHZ^{\epsilon}_N}} 
    = N_{\epsilon} ( \ket{0}^{\otimes N} + \ket{\epsilon}^{\otimes N}  ),
\end{equation} 
where the normalization is given by $N_{\epsilon}=1/\sqrt{2(1+\cos^N(\epsilon)}$. For $\epsilon=\pi/2$, the GHZ state is recovered (Cat A). For $\epsilon=0$, a trivial microscopic product state arises: $\ket{0}^{\otimes N}$, which is analogous to Cat B (one can write $\ket{0} =1/\sqrt{2} (\ket{+}+\ket{-})$). Let us look at the case where $\epsilon$ is very small, i.e. $\epsilon \rightarrow 0$. In this case, the state of the individual particle $\ket{\epsilon}$ is very close to $\ket{0}$, so their overlap is big: $|\langle 0 |\epsilon\rangle|^2=\cos^2(\epsilon) \approx 1-\epsilon^2 \approx 1.$ However, when we consider $N \gg 1$ particles and look at the overlap between the two branches of the generalized GHZ state, that is, between $\ket{0}^{\otimes N}$ and $\ket{\epsilon}^{\otimes N}$, we see that the branches are almost orthogonal (zero overlap), since $\ket{\epsilon}^{\otimes N} \approx \cos^N(\epsilon) \ket{0}^{\otimes N}$ and thus,
\begin{equation}
    |\langle 0 |^{\otimes N} |\epsilon\rangle^{\otimes N}|^2 \approx \cos^{2N}(\epsilon) \approx (1-\epsilon^2)^N \approx e^{-N\epsilon^2}, 
\end{equation}
which tends to zero for $\epsilon > 1/\sqrt{N}.$ Effectively, it is as if $\ket{\epsilon}^{\otimes N} \approx \ket{1}^{\otimes N}$, which is what a GHZ state has as second branch of the superposition. Yet, even though the generalized GHZ state is a superposition of two (almost) orthogonal branches with a large number of particles each, the authors in~\cite{frowis2012measures} showed that the macroscopicity of this state is significantly smaller than that of a GHZ state. Precisely, it is $N_{\rm eff} \approx N \epsilon^2$ (compared to $N_{\rm eff} = N$ for the GHZ state). This result comes from the computation of the variance with respect to an optimized observable $D_{\rm opt}$, giving~\cite{frowis2012measures} $(\Delta D_{\rm opt})^2 \approx N^2 \sin^2(\epsilon)$. Using Eq.~\eqref{Neff} and $\sin^2(\epsilon) \approx \epsilon^2$, we conclude that the effective system size for small $\epsilon$ is $N_{\rm eff} \approx N \epsilon^2$. As an example, this result implies that a generalized GHZ state with $N=10^7$ particles and $\epsilon = 10^{-3}$ has an effective size of an ideal GHZ state with $10$ particles.

\subsection{Cat state from a superposition of two coherent states}

In this section, we discuss the oldest and best studied example for the transition between quantum and classical physics. Coherent states have been introduced by Schr\"odinger in 1925 in order to discuss the transition between classical and quantum harmonic oscillators in the phase space picture. The coherent state can be viewed as the quantum oscillator which is closest to its classical counterpart. They are often used in the context of quantum optics to describe light, e.g. the coherent state of a laser.

The Hamiltonian defining the harmonic oscillator of a particle with mass $m$ that oscillates with frequency $\omega$ is given by
\begin{equation}
    \hat{H} = \frac{\hat{p}^2}{2m} + \frac{1}{2} m \omega^2 \hat{x}^2 = \hbar \omega (\hat{a}^\dagger \hat{a} + \frac{1}{2}) = \hbar \omega (\hat{N}+\frac{1}{2}),
\end{equation}
where $\hat{p}$ is the momentum operator, $\hat{x}$ is the position operator, $\hat a^\dagger$ and $\hat a$ are the so-called creation and annihilation operators, respectively, and $\hat{N}$ is the number operator. The energy eigenstates $\ket{n}$ are defined as $\hat H \ket{n} = E_n \ket{n}$, with $E_n = \hbar \omega (n+1/2)$ and $n=0,1,2,...$; which are also eigenstates of the number operator, i.e., $\hat{N}\ket{n} = n\ket{n}.$

Coherent states are defined as the eigenstates of the annihilation operator as
\begin{equation}
    \hat a \ket{\alpha} = \alpha \ket{\alpha},
\end{equation}
where $\alpha$ is a complex number, i.e. $\alpha = |\alpha| e^{i \phi}$,  because $\hat{a}$ is not hermitian. It is possible to obtain an explicit expression for the coherent state in terms of the energy eigenstates $\ket{n}$ as 
\bea
|\alpha\rangle = e^{-|\alpha|^2/2} \sum_{n=0}^\infty \frac{\alpha^n}{\sqrt{n!}} | n \rangle.
\eea
We can also define the coherent state as $\ket{\alpha}=D(\alpha)|0\rangle$ with the displacement operator
\bea
D(\alpha)=\exp (\alpha \hat a^\dagger - \alpha^* \hat a).
\eea

\begin{figure}
\begin{center}
\includegraphics[scale=1]{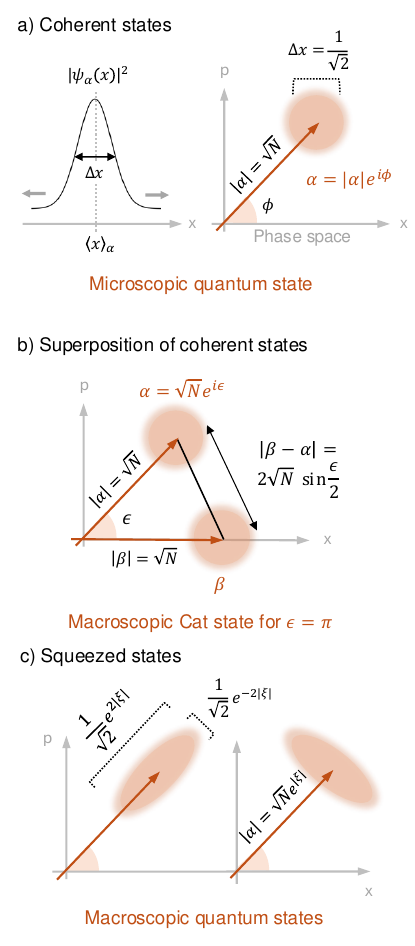}
\end{center}
\caption{a, left) A coherent state represented as a gaussian with expected value $\langle x\rangle_\alpha$, which oscillates periodically (grey arrows) in the position representation. a, right) In phase space, a coherent state is visualized as a complex number $\alpha = |\alpha|e^{i\phi}$, which has minimal variance both in momentum and position and thus is a non-macroscopic quantum state. b) A superposition of two coherent states $\propto \ket{\alpha} + \ket{\beta}$ can lead to a macroscopic quantum state for $\epsilon=\pi$ (see text for details). c) Squeezing of the coherent state either in phase or in particle number can lead to a macroscopic quantum state.} \label{fig:squeeze}
\end{figure}

Its wave function expressed in terms of the position is $\ket{\alpha} = \int dx \, \psi_\alpha(x) \ket{x}$, whose coefficients $\psi_\alpha(x)$ in turn give the probability density function of the particle position as
\begin{equation}
    |\psi_\alpha(x)|^2 = \frac{1}{\sqrt{2\pi} \Delta x_\alpha} e^{-\frac{1}{2} \left( \frac{x-\langle x \rangle_\alpha}{\Delta x_\alpha}  \right)^2},
\end{equation}
where we see that it is a Gaussian centered at $\langle x \rangle_\alpha$ with width $\Delta x_\alpha$. In the classical harmonic oscillator, the particle's position oscillates back and forth following a sinusoidal function. Analogously, in the quantum case, the \textit{expected value} of the coherent state's position oscillates back and forth in the position representation (see Fig.~\ref{fig:squeeze}(a, left)) following $\langle x \rangle_\alpha (t) \propto \cos{(\omega t - \varphi)}$, with an arbitrary $\varphi$ given by the initial conditions. The classical harmonic oscillator can be represented in phase space (momentum as a function of position), as a point that moves around a circumference of radius 1. Analogously, we can also represent coherent states in terms of its momentum and position as we see in Fig.~\ref{fig:squeeze}(a, right), where we also depict its relations to the real and imaginary parts of the complex number $\alpha$ that defines the state:  
\begin{align}
    &\langle x \rangle_\alpha = \sqrt{\frac{2\hbar}{m\omega}} \textrm{Re}(\alpha), \nonumber \\
    &\langle p \rangle_\alpha = \sqrt{2m\hbar \omega} \textrm{Im}(\alpha).
\end{align}

One important feature of coherent states is that they have minimal uncertainty in momentum and position. Heisenberg's uncertainty relation shows that $\Delta x \Delta p \geq \hbar /2$. Coherent states achieve the lower bound $\Delta x \Delta p = \hbar /2$. 

The coherent state $|\alpha= 0 \rangle$ has minimal variance also in fluctuations in phase and in number of particles in Fock space, that is, $\Delta \phi \Delta n = 1/2$. Thus, a single coherent state is a microscopic quantum state. We can show this by determining the effective size $N_{\rm eff}$ of the system, for which we first select an operator $D$ and calculate its variance $(\Delta D)^2$ (see Eq.~\eqref{Neff}). In this case, the operator $D$ we consider is the number operator $\hat{N}$. The expectation value of the number operator (average number of particles, e.g. photons, which we denote as $N$) can be related to the modulus of $\alpha$ as
\bea
N := \langle \hat{N}\rangle =\langle \alpha | \hat{a}^{\dagger} \hat{a}|\alpha\rangle =   |\alpha|^2.
\eea
Using $[\hat{a}, \hat{a}^{\dagger}]=\hat{a} \hat{a}^{\dagger}-\hat{a}^{\dagger}\hat{a} = 1$, we further get
\begin{align}
\langle \hat{N^2}\rangle &=\langle \alpha | \hat{a}^{\dagger} \hat{a} \hat{a}^{\dagger} \hat{a}|\alpha\rangle
\nonumber\\
\nonumber
&= \langle \alpha | \hat{a}^{\dagger} \hat{a}^{\dagger} \hat{a} \hat{a}|\alpha\rangle + \langle \alpha | \hat{a}^{\dagger} \hat{a}|\alpha\rangle \\ &=|\alpha|^4 + |\alpha|^2,
\end{align}
so the variance of the number operator is given by $(\Delta D)^2= \langle \hat{N^2}\rangle -\langle \hat{N}\rangle ^2= |\alpha|^2 =N$. Thus, the effective system size is $N_{\rm eff} \propto (\Delta D)^2/N = {\cal O}(1)$, indicating that the coherent state indeed is a microscopic state.

There are two possibilities to construct macroscopic quantum states using coherent states: superpositions of two coherent states, and squeezing of a given coherent state, see Fig.~\ref{fig:squeeze}(b,c). We first discuss superpositions of coherent states, also known as cat states, defined as
\bea
\ket{\textrm{CAT}} ={\cal N}_{\alpha, \beta}(\ket {\alpha}  + \ket {\beta}),
\eea
where ${\cal N}_{\alpha, \beta}$ is the normalization of the coherent state, given by 
\bea
{\cal N}_{\alpha, \beta}= \frac{1}{\sqrt{2 + \bra{\alpha}\beta\rangle + \bra{\beta}\alpha\rangle}}.
\eea
Since the number operator is not sensitive to the phase (in other words, the number of quanta is always positive), in order to determine the macroscopicity of the cat-state, it is convenient to rescale the cat state to 
\bea
\ket{{\rm{cat}}} ={\cal N}_{0, \beta-\alpha}(\ket {0}  + \ket {\beta-\alpha}) 
\eea
with the normalization given by
\begin{equation}\label{eq:norm_sup_coherent}
    {\cal N}_{0, \beta-\alpha} = \frac{1}{\sqrt{2+2\, e^{-\frac{1}{2}|\beta - \alpha|^2}}},
\end{equation}
where we have used
\begin{equation}
    \bra{0} \beta - \alpha \rangle = \bra{\beta-\alpha} 0 \rangle = e^{-\frac{1}{2}(|0|^2 + |\beta-\alpha|^2)} \cdot e^{0 \cdot (\beta-\alpha)},
\end{equation}
(coherent states are not orthogonal to each other).
Similarly as above, we calculate the number operator variance $(\Delta D)^2$ and find
\begin{align}\label{eq:variance_sup_coherent}
    (\Delta D)^2 &= \langle \hat{N^2}\rangle_{cat} -\langle \hat{N}\rangle_{cat} ^2 \nonumber\\
    &=(|{\cal N}_{0, \beta-\alpha}|^2-|{\cal N}_{0, \beta-\alpha}|^4)\cdot  |\beta - \alpha|^4 \nonumber\\
    &+ |{\cal N}_{0, \beta-\alpha}|^2 \cdot |\beta - \alpha|^2.
\end{align}
For the extreme case $\beta = -\alpha  \equiv - \sqrt{N}$, the overlap between the two coherent states is exponentially suppressed, as we can approximate the normalization in Eq.~\eqref{eq:norm_sup_coherent} as ${\cal N}_{0, \beta-\alpha} \approx 1/\sqrt{2}$, since $e^{-\frac{1}{2}|\beta - \alpha|^2} \rightarrow 0$ (for large $N$). With this approximation, we then use Eq.~\eqref{eq:variance_sup_coherent} to get the effective size
\begin{align}
    N_{\rm{eff}} &= \frac{\frac{1}{4}|2 \alpha|^4 + \frac{1}{2}|2 \alpha|^2}{N} = 4 N + 2 = \mathcal{O}(N),
\end{align}
where we have considered $|\alpha|^2=N$ (see Fig.~\ref{fig:squeeze}(b)).

On the other hand, if the distance $|\beta - \alpha| = 2 \sqrt{N} \sin(\frac{\epsilon}{2})$ is small (see Fig.~\ref{fig:squeeze}(b) for trigonometric considerations), the overlap between the coherent states becomes very large and the normalization ${\cal N}_{0, \beta-\alpha}$ becomes approximately $1/2$, while the expectation value for the particle number remains $N (1 + {\cal O} (\epsilon^2))\simeq N$. In this limit, the effective number of  particles involved in the macroscopic state shrinks down to ${\cal O}(1)$ like $N \epsilon^4$.

\subsection{Transition to squeezed states}

As shown in Fig.~\ref{fig:squeeze}(c), there is yet another possibility to find a transition from a microscopic to a macroscopic state starting from a coherent state. Rather than considering the superposition of two coherent states at angular distance $\epsilon$, we start with a single coherent state.   In the squeezed state, the idea is to increase the fluctuations of one of two incompatible observables, at the expense of the fluctuations of the other one, so that Heisenberg's uncertainty relation still holds, but the variance of one observable is increased, and so does the effective size of the state. For example, either fluctuations in photon number are increased by $e^{+2 |\xi|}$ and decreased in phase by $e^{-2|\xi|}$, or vice versa, with constant product $e^{+2 |\xi|} e^{-2 |\xi|} =1$. Thus, there is no natural decomposition of the state as a `cat state' in the form $\ket{\phi_1}+\ket{\phi_2}$ available. However, also in this case, a transition from a microscopic quantum state to a macroscopic quantum state can be deduced using the quantum Fisher information to determine the effective size of the system. 

Formally, squeezing can be done by introducing the squeezing operator
\bea
S(\xi)=\exp ( -\frac{1}{2}[ \xi^* (\hat a)^2-\xi ({\hat a}^\dagger)^2]).
\eea
Then, using the  displacement operator
\bea
D(\alpha)=\exp (\alpha \hat a^\dagger - \alpha^* \hat a),
\eea
we define the squeezed state as
\bea
\ket{\xi, \alpha} = 
S(\xi) D(\alpha) \ket{0}.
\eea

In order to calculate expectation values, and to keep expressions simple, we only consider real $\xi$ and find
\bea
S(\xi)^{\dagger} a  S(\xi)  = a \cosh{\xi} + a^{\dagger} \sinh{\xi} \equiv a_s,
\eea
and in the same manner we introduce $a^{\dagger}_s = a^{\dagger} \cosh{\xi} + a \sinh{\xi}$ for real $\xi$. In general, squeezing can be understood as an $SU(1, 1)$-transformation of the creation and annihilation operators. Note that $[a, a^{\dagger}] = [a_s, a_s^{\dagger}] = 1 $.
For real $\alpha$, the expectation value of  $a_s $ scales like $\alpha e^{\xi}$,  since
\begin{align}
\bra{\xi, \alpha} a  \ket{\xi, \alpha} &=
\bra{\alpha} S(\xi)^{\dagger} a  S(\xi) \ket{\alpha} 
\nonumber\\ \nonumber
&=\bra{\alpha} a_s  \ket{\alpha} \\
& =  \alpha \cosh{\xi} + \alpha^* \sinh{\xi}.
\end{align}

For real $\alpha, \xi$, it is easy to see that $\bra{\xi, \alpha} a  \ket{\xi, \alpha} = |\alpha|e^{\xi}$.  Generalization to complex  $\alpha, \xi$, leads to the same result plus a phase factor, see Fig. (\ref{fig:squeeze}) c.  Next, we calculate the variance, defined as

\begin{align}\label{eq:variance_sup_coherent}
   (\Delta D)^2  &= \bra{\xi, \alpha}  \hat{N}^2\ket{\xi, \alpha}  -  \bra{\xi, \alpha}  \hat{N}\ket{\xi, \alpha} ^2 \nonumber\\
&=
\bra{\xi, \alpha}  a^{\dagger} a a^{\dagger} a  \ket{\xi,\alpha}  -  \bra{\xi, \alpha}  a^{\dagger} a\ket{\xi, \alpha} ^2 \nonumber\\
&=
\bra{\alpha} a_s^{\dagger} a_s \ket{\alpha}. \ \ \ \ \ \ \ \ \ \ \ \ \ \ \ \ \ \  \ \ \ \ \ \ \ \ \ \ \ \ \ \ \ 
    \end{align}
    We used $[a, a^{\dagger}] = [a_s, a_s^{\dagger}] = 1 $ and unitarity of the squeezing operator, $S(\xi)^{\dagger} S(\xi) = \mathbb{1}$, and thus $S(\xi)^{\dagger} a a S(\xi) = S(\xi)^{\dagger} a S(\xi)^{\dagger} S(\xi)a S(\xi) = a_s a_s$, etc.
For real $\alpha, \xi$, we obtain as result  $|\alpha|^2e^{2 \xi}$  and
$N_{\rm eff} = \frac{(\Delta D)^2}{|\alpha|^2} = e^{2 |\xi|}$, which eventually becomes macroscopic for $|\xi| \simeq \ln (\sqrt{N}) = \ln \alpha$.  Note that we divide by the factor $|\alpha|^2$  which represents the number of particles $N$ of the original coherent state without squeezing. It can be shown that the result $N_{\rm eff} = e^{2 |\xi|}$ holds also in general. For more details, we refer to \cite{Frowis2017} and references therein.

\subsection{Dicke states}

In this section, we return to qubits to discuss another important class of quantum states, the so-called Dicke states, defined as
\bea
\label{Dicke}
| D_N^{(k)} \rangle = 
{N \choose k}^{-1/2}  \sum_{\Pi} \Pi_k (| 0^{\otimes k} 1^{\otimes N-k} \rangle ).
\eea
Here $\Pi_k$ describes all ${N \choose k}$ permutations of $k$ states $|0\rangle$ with $N-k$ states $|1\rangle$.  We may ask whether the Dicke state is ``macroscopic'' or not. In order to answer this question, we have to determine the effective size $N_{\rm eff}$, that is, the size of the ``super-cells'' within our cat. In appendix  \ref{Dicke}, we show that $N_{\rm eff} = \frac{1}{N}(2 k (N-k)) + 1$. That is, Dicke states are not macroscopic for $k=1$, but indeed become macroscopic for $k=N/2$. Intuitively speaking, fluctuations are obviously zero for $k=0$, and gradually increase with increasing number of possibilities to ``flip'' between dead and alive cells, that is, between $0$ and $1$. As $N_{\rm eff}$ can be viewed as a measure of these possibilities for flips, the maximal effective size $N_{\rm eff} = N/2 +1$ is reached for $k=N/2$. The simplest and most important Dicke state is the so-called W-state
\bea
\label{Dicke}
| W \rangle = \frac{1}{\sqrt{3}}
(|001\rangle+|010\rangle+|100\rangle)
\eea
which is the second type of 3-particle entanglement besides the GHZ-state $| \textrm{GHZ} \rangle =\frac{1}{\sqrt{2}}( 
|000\rangle+|111\rangle)$ for the case of $N=3$ qubits. While the GHZ-state for $N \gg 1$ is a macroscopic quantum state (our cat A),  the generalized W-state, that is, the Dicke state, is only macroscopic if the number of ``dead'' and ``alive'' cells are about the same size, that is, for $k \simeq N/2$.

%Table of examples, discuss whether these are  "macroscopic quantum effects" or "accumulated microscopic effects"

%\begin{table*}
%\begin{tabular}[h]{c|c}
%Example  & Macroscopic or not? \\
%\hline
%Product of Bell-state:   & accumulated microscopic quantum system  \\
%\hline
%Interference of two BEC  & xxx  \\
%\hline
%Double slit with bucky balls & xxx  \\
%\hline
%Ferromagnetism & under debate  \\
%\hline
%Superposition of currents in SQUID & under debate  \\
%\hline
%\end{tabular}
%\begin{caption}{ "macroscopic quantum" or "accumulated microscopic" quantum system?}
%\end{caption}
%\end{table*}

\section{Application of macroscopic quantum effects to quantum metrology}\label{sec:macrostates_and_metrology}

\subsection{Increased phase sensitivity of macroscopic quantum states}

Metrology is the science of measurement, where experiments are typically designed to estimate the value of a certain parameter as precisely as possible. The main idea is to subject the system to a time  evolution involving the parameter $\varphi$ that is to be estimated. For example, if one wants to measure the value of a magnetic field, one can make a system feel the magnetic field and then compare the final state of the system to the initial one. The more the state changes due to the interaction with the magnetic field, the easier it would be to estimate its value from measurements on the final state. This idea is at the core of statistical measures like the Fisher Information $F(\varphi)$, which mathematically quantifies the change of the state with respect to the parameter $\varphi$ (see also Appendix~\ref{app:QFI_pure_states}) as
\begin{equation}\label{eq:FI_metrology}
    F(\varphi) = \sum_i p_i(\varphi) \left[ \frac{d}{d\varphi} \ln{p_i(\varphi)}\right]^2,
\end{equation}
where $\varphi$ is the parameter we want to estimate. As we explained above, the final state that we are measuring depends on the parameter, so the outcomes of the measurement will do so, too. These outcomes are distributed following a probability density function that we denote as $p(\varphi)$ here. Hence, $p_i(\varphi)$ is the probability of measuring outcome $i$. The quantum version of the Fisher Information, which we denote as $\mathcal{F}(\varphi)$, is obtained by maximizing over all possible measurements of the observable in question (e.g. the magnetic field). More technically, this maximization occurs over all possible sets of Positive Operator-Valued Measure (POVMs) that define the measurement of the observable. 

The more a state $|\psi \rangle$ changes w.r.t. the parameter $\varphi$, the lower the error $\Delta \varphi$ in the estimation, which is formally defined by the so-called Cram\'er-Rao bound
\begin{equation} \label{eq:cramer-rao_bound}
    \Delta \varphi \geq \frac{1}{\sqrt{\mathcal{F}(\varphi)}},
\end{equation}
which is inversely proportional to the square root of  QFI. By looking at this expression, we can see the relation between macroscopic quantum states and metrology. As we saw in Section~\ref{sec:QFI}, macroscopic quantum states have larger QFI ($\mathcal{F}=O(N^2)$) than other states that are not truly macroscopically quantum ($\mathcal{F}=O(N)$). Therefore, they can be used as a resource to enhance the precision of the parameter estimation. Such enhancement can be quantified by the Cram\'er-Rao bound~\eqref{eq:cramer-rao_bound}, which scales as $1/N$ (also known as Heisenberg scaling) for macroscopically quantum states, in contrast to the $1/\sqrt{N}$ scaling of classical and non-macroscopic quantum states. Physically speaking, the  $1/\sqrt{N}$ scaling corresponds to higher accuracy due to independent, repeated measurements. The enhanced accuracy factor $1/N$ is due to phase accumulation upon entanglement, as we will make more precise in what follows.

\begin{figure}
\centering
\includegraphics[width=3.3in]{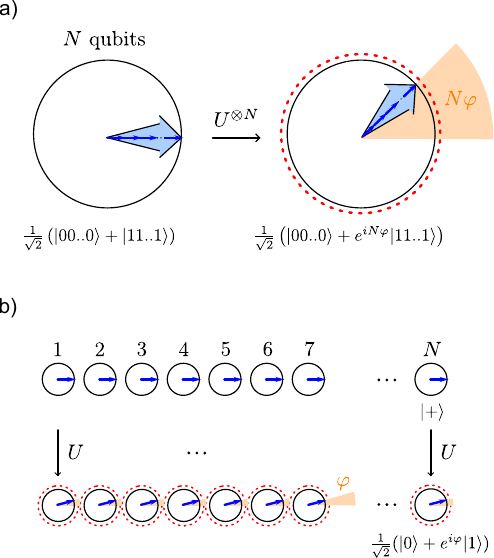}
\caption{A system of $N$ qubits in the quantum states (a) $\ket{\textrm{GHZ}}$ and (b) $\ket{+}^{\otimes N}$ evolves under the action of a unitary evolution $U^{\otimes N}$, where the single qubit unitary is defined as $U = \ketbra{0}{0} + e^{i \varphi} \ketbra{1}{1}$. Qubits are depicted as vectors in the so-called Bloch sphere (see~\cite{dur2014visualization}) where a unitary evolution is a rotation of the vector with an angle $\varphi$. (a) Entangled qubits in the GHZ state rotate as a collective and behave as a ``super qubit'' that rotates an angle $N$ times bigger than the angle $\varphi$ that defines the unitary evolution of the individual qubits. (b) Qubits in the state $\ket{+}^{\otimes N}$ are not entangled and each qubit rotates independently an angle $\varphi$. Dashed red circles depict a measurement process, carried out after the evolution.}\label{fig:visual_QFI_metrology}
\end{figure}

Fig.~\ref{fig:visual_QFI_metrology} shows the difference between estimating a parameter with $N$ qubits in state $\ket{+}^{\otimes N}$ and $N$ qubits in the GHZ state. We have chosen as an example a unitary evolution of the form
\begin{equation}
    U(\varphi) = \begin{pmatrix}
        1 & 0 \\ 0 & e^{i\varphi}
    \end{pmatrix}
\end{equation}
which introduces a phase $e^{i\varphi}$ if the state is $\ket{1}$ and leaves the state $\ket{0}$ unchanged. As a global phase is irrelevant, we may have Larmor-precession in an external magnetic filed in mind as explicit application, leading to this unitary time evolution. When applied to e.g. a superposition state $\ket{+}$, it gives
\begin{equation}
     U(\varphi) \ket{+} = \frac{1}{\sqrt{2}} \left( \ket{0} + e^{i\varphi} \ket{1} \right). \label{eq:Uplus}
\end{equation}

Each particle evolves under the action of this unitary, so the global unitary acting on the entire state of $N$ particles is $U^{\otimes N}$. First, we apply this unitary to the GHZ state:
\begin{align}\label{eq:evolved_GHZ_metrology}
    U^{\otimes N} \ket{\textrm{GHZ}} &= \frac{1}{\sqrt{2}} \left( U^{\otimes N} \ket{00..0} + U^{\otimes N} \ket{11..1} \right) \nonumber \\
    &= \frac{1}{\sqrt{2}} \left( \ket{00..0} + e^{i\varphi}\cdot e^{i\varphi} \cdot e^{i\varphi}\cdot...\cdot e^{i\varphi} \ket{11..1} \right) \nonumber \\
    &= \frac{1}{\sqrt{2}} \left( \ket{00..0} + e^{iN\varphi} \ket{11..1} \right),
\end{align}
where we see that the resulting state's form is analogous to the one in Eq.~\eqref{eq:Uplus} but with $N$ particles instead of one and a phase that is $N$ times enhanced. Due to entanglement, all particles are merged an feel the phase as a collective, which can be visualized as in Fig.~\ref{fig:visual_QFI_metrology}a, where a ``super qubit'' composed of many qubits rotates as a whole with an amplified rotation angle $N \varphi$. In this simple example, the estimation error can be computed. As mentioned above, once the initial state has been subjected to the $\varphi$-dependent evolution, the probability of obtaining a certain outcome will depend on $\varphi$, so it can be used as an estimator for the parameter. Precisely, we can choose the probability of obtaining the initial state as estimator (also known as "fidelity"), which reads
\begin{equation}\label{eq:estimator_GHZ_metrology}
    p(\varphi) = |\langle \psi_{\textrm{init}} | \psi_{\textrm{fin}}\rangle |^2 = \left|\frac{1}{2} (1+e^{iN\varphi)} \right|^2 = \cos^2{\frac{N\varphi}{2}}.
\end{equation}
We can obtain the uncertainty $\Delta \varphi$ by propagating the uncertainty in $p(\varphi)$ as $\Delta \varphi = \Delta p(\varphi) / (|\partial p (\varphi)/ \partial \varphi|)$, where 
\begin{equation}
    \Delta^2 p(\varphi) = \langle P^2 \rangle - \langle P \rangle^2 = p(\varphi)-p^2(\varphi),
\end{equation}
where $P=\ketbra{\psi_{\textrm{init}}}{\psi_{\textrm{init}}}$ is the projector onto the initial state that defines the measurement, and $\langle P \rangle = \bra{\psi_{\textrm{fin}}} P \ket{\psi_{\textrm{fin}}} = p(\varphi)$. Making use of Eq.~\eqref{eq:estimator_GHZ_metrology}, we get an error
\begin{equation}
    \Delta p(\varphi)= \cos{\frac{N\varphi}{2}}\sin{\frac{N\varphi}{2}},
\end{equation}
which is $\frac{1}{N} \cdot (|\partial p (\varphi)/ \partial \varphi|)$, so the error in the parameter estimation is $\Delta \varphi = 1/N$.

We compare this case to our second example, the state $\ket{+}^{\otimes N}$, which evolves as
\begin{align}
     U^{\otimes N} \ket{+}^{\otimes N} &= U \ket{+} \otimes ... \otimes  U \ket{+} \nonumber \\
     &= \left( \frac{1}{\sqrt{2}} ( \ket{0} + e^{i\varphi} \ket{1}) \right)^{\otimes N}.
\end{align}
As we see in Fig.~\ref{fig:visual_QFI_metrology}b, each qubit rotates independently with an angle $\varphi$. This process can also be thought of as making one qubit undergo the evolution and then measuring it. The computation of the error $\Delta \varphi$ is analogous to the one above for the GHZ state, now with $p(\varphi) = |\langle \psi_{\textrm{init}} | \psi_{\textrm{fin}}\rangle |^2 = \cos^2{\frac{\varphi}{2}}$. In this case, $\Delta \varphi = 1$. If one repeats the experiments $N$ times (with $N$ qubits), the estimation error decreases by a factor $1/\sqrt{N}$, i.e. $\Delta \varphi = 1/\sqrt{N}$. As a conclusion, for macroscopic states with $N_{\rm eff} = N$, the estimation error is reduced by a factor $1/\sqrt{N_{\rm eff}}$ in comparison to an accumulation of microscopic quantum states with $N_{\rm eff} = 1$.

In Appendix~\ref{app:cramer-rao_bound}, we compute the Cram\'er-Rao bound for these two examples by means of Eq.~\eqref{eq:FI_metrology}, which illustrates how the aforementioned ideas for parameter estimation unfold in a mathematical way. We refer the reader to~\cite{giovannetti2004quantum} for more examples with different unitary evolutions and metrology setups.

\subsection{Reduced decoherence time of macroscopic quantum states}

In the previous section, we have shown that the higher sensitivity of phase factors in macroscopic quantum states decreases the measurement uncertainty by a factor of order $1/\sqrt{N_{\rm eff}}$. However, it unavoidably also increases decoherence. In Fig. \ref{fig:visual_QFI_metrology}, we display a coherent state evolution assuming that the interaction with the noisy environment is suppressed. However,  this is very difficult to achieve, and in what follows, we will discuss the effect of incoherent interaction. In \cite{Deco2018}, a simple decoherence model has been introduced to show that the time evolution of a single qubit ($N=1$), in a state described by the density matrix $\rho(0)=|+ \rangle \langle +|$, that is coupled to a noisy environment with $M$ qubits is obtained by ``tracing out the environment''. In other words, we reduce the Hilbert space of the system S (N qubits) and environment E (M qubits) by calculating the partial trace ${\rm tr}_E \rho_{S \times E}$. As shown in \cite{Deco2018}, the result is (for $N=1$)
\begin{eqnarray}
 \rho(t)= \frac{1}{2}|0 \rangle \langle 0| + \frac{1}{2}|1 \rangle \langle 1| +
 \\ \nonumber
 \frac{1}{2} \prod_{k=1}^M \cos(\omega_k) |1 \rangle \langle 0| + \frac{1}{2}  \prod_{k=1}^M \cos(\omega_k)|0 \rangle \langle 1|.
  \\ \nonumber
\end{eqnarray}
Here, $\hbar \omega_k$ is the interaction energy of the $k$-th qubit interaction with the system qubits via the Hamiltonian $\hbar \omega_k (\sigma_z^k \otimes \sigma_z)$. In this simple model, using the approximation 
\begin{eqnarray}
\prod_{k=1}^M \cos(\omega_k) &\simeq& 
\\ \nonumber
\exp(\sum_k \log (1 -  1/2 (\omega_i t)^2)) 
&\simeq& \exp(- t^2/\tau_{dec}^2),
\end{eqnarray}
the decoherence time is found to be $\tau_{\rm dec} = \sqrt{2/ \langle (\Delta \omega)^2 \rangle}$. The broader the frequency spectrum of the noisy environment, the faster the decoherence. 

Using the same model, if we now consider our Cat B with density matrix $\rho_B(0)=|+ \rangle^{\otimes N} \langle +|^{\otimes N}$, we find that the decoherence time of any off-diagonal element reads
\begin{eqnarray}
(\prod_{k=1}^M \cos(\omega_k))^N &\simeq& 
\exp(- N t^2/\tau_{dec}^2),
\end{eqnarray}
leading to faster decoherence than the single qubit case presented above. Precisely, the decoherence time is reduced by a factor $\sqrt{N}$, $\tau^{N=1}_{\rm dec} \rightarrow \tau^{N}_{\rm dec}\equiv \tau_{\rm dec}/\sqrt{N}$, when compared to the single-qubit decoherence.

On the other hand, if we consider our Cat A with density matrix $\rho_A(0)=| \rm{GHZ_N} \rangle\langle \rm{GHZ_N}|$ we find that its decoherence time is
\begin{eqnarray}
(\prod_{k=1}^M \cos(N \omega_k)) &\simeq& 
 \exp(- N^2 t^2/\tau_{dec}^2),
\end{eqnarray}
which is even faster. As shown in Eq.~\eqref{eq:evolved_GHZ_metrology}, the accumulated phase factor $N \varphi$ leads to higher precision in measurements for macroscopic quantum states with $N_{\rm eff}=N$ by a factor of $1/\sqrt{N_{\rm eff}}$.  Concerning decoherence, comparing cat A (macroscopic quantum state with $N_{\rm eff}=N$) with cat B (microscopic), the decoherence time of cat A is $1/\sqrt{N_{\rm eff}}$ times smaller than that of cat B. This is the price to pay for the higher phase sensitivity of macroscopic quantum states shown in Fig \ref{fig:visual_QFI_metrology}, precisely reflecting the transition in sensitivity from the classical limit (micrscopic quantum system) to the Heisenberg limit (macroscopic quantum system). This is also the reason why it is notoriously difficult to prepare GHZ states for a large number $N$ of qubits.

% How to prepare Dicke states

% Aspelmeyer, Obertaler, etc

% Start with limits of resolution

% \begin{itemize}
% \item{Experiments with entangled ion traps (B-field, Gravitation)}
% \item{Examples with GHZ-like states, and $N/2$-Dicke states, and...}
% \end{itemize}

% Scaling argument using variance and $N_{\rm eff}$.

% Motivate why variance is a good measure for this sensitivity, Introduce variance $\langle \Delta A^2\rangle$ of local operator $A = \sum_{l=1}^N A^{l}$. Show that variance for GHZ state is maximal $\propto N^2$ for $A = \sum_{l=1}^N \sigma_z^{l}$ - Motivate definition of 

% \bea
% N_{\rm eff} = \frac{1}{N}{\rm Max}_{A: \rm{local}} \langle \Delta A^2\rangle
% \eea

% Discuss definition of "Macroscopic" quantum effect: Formally $N_{\rm eff} \propto N$ in contrast to $N_{\rm eff} \propto o(1)$.

\section{Wanted: Schrödinger cat - dead and alive}
\label{sec:DeadAlive}

Various systems have been discussed to generate macroscopic quantum superposition states in the laboratory, and show that these puzzling quantum effects do not only exist on paper, but can be observed in experiments. This includes experiments with optical systems, more precisely optical photons as well as microwave photons. Demonstrations have also been performed using spin systems, both with individual control such as with trapped ions, and with collective control, e.g. using Bose Einstein Condensates. Finally, superpositions of massive systems have been experimentally demonstrated.

With trapped ion systems, where electronic states represent qubits, there exists a clean and high-fidelity implementation of GHZ states $|0\rangle^{\otimes n}+|1\rangle^{\otimes n}$, for $n$ up to 14~\cite{monz2011fourteenqubit}. Actually, a staggered basis $|01\rangle^{\otimes n/2}+|10\rangle^{\otimes n/2}$ was used, since such a state is not affected by global field fluctuations, which constitute one of the main noise sources in these set-ups. Due to some noise effects and imperfections, the effective size that can be associated with the generated states is about 11, slightly below the actual particle number~\cite{Frowis2017}. The assessment of an effective size is thereby done using the usefulness of the quantum state for metrology~\cite{Frowis2017}, utilizing the connection of macroscopicity with the quantum Fisher information~\cite{frowis2012measures}.

In other set-ups, much larger number of spins were involved. In fact, spin squeezed states of several thousand atoms have been reported in different experiments, with even up to half a million atoms in~\cite{hosten2016measurement}. However, this does not necessarily mean that the states are as macroscopic as a GHZ state of this size, since they are of a different form. They reach effective sizes of about 71~\cite{Frowis2017}. Actual advantages in quantum sensing have also been demonstrated using trapped ions, where a variational approach was employed to generate useful states, and demonstrate a quantum advantage with up to 26 ions~\cite{marciniak2022optimal}. Given the close connection between usefulness for sensing and macroscopicity~\cite{frowis2012measures}, these states can also be considered to be macroscopic. Notice that the utilized states were not of GHZ form, since these states are not only very good to sense magnetic fields as outlined in Sec.~\ref{sec:macrostates_and_metrology}, but are also very fragile and sensitive to noise, errors and imperfections.

Squeezing has also been successfully applied in photonic experiments to generate macroscopic states of optical photons~\cite{vahlbruch2016detection}, where the effective size reached in this experiment is 32. Squeezed states of light are also used in gravitational wave detectors such as LIGO to boost the achievable precision. 

Another class of states that have been demonstrated experimentally are the so-called cat states, which constitute a coherent superposition of coherent states $|\alpha\rangle$ and $|-\alpha\rangle$. Coherent states are viewed as classical states of a harmonic oscillator, and one has hence a superposition of two distinct classical states as is the spirit of Schrödinger cat states. Such states have been realized using microwave photons~\cite{deleglise2008reconstruction, vlastakis2013deterministically, wang2016schrodinger}, and with trapped ions~\cite{kienzler2016observation}, leading to effective sizes of 10 (20) for the microwave experiments, and 49 for the trapped ion realization. The latter is actually an experiment with massive systems, where a coherent superposition between two objects with a spatial separation a dozen times larger than the local position fluctuations was generated. Even larger spatial separations have been achieved with Bose Einstein condensates, where about $10^5$ Rubidium atoms where separated by more than half a meter~\cite{kovachy2015quantum}. Experiments are also ongoing with opto-mechanical systems~\cite{aspelmeyer2014cavity,barzanjeh2022optomechanics} where e.g. massive optical mirrors are put in coherent superposition. Finally, in matter wave interferometry~\cite{gerlich2011quantum}, wave properties of systems of up to several thousand atomic mass units have been observed in double slit experiments.

For further details, we refer interested readers to the review~\cite{Frowis2017}.

\section{Discussion}
The transition from the microscopic to the macroscopic world has been a subject of debate since Schr\"odinger devised his famous thought experiment. His cat was clearly a classical macroscopic system that ended up being in a paradoxical situation when it got coupled to a microscopic quantum system that was in a superposition state. With his experiment, Schr\"odinger wanted to show the world the paradoxes that one gets into when trying to extend the quantum theory and its Copenhagen interpretation to our everyday, macroscopic world. 

Based on this seminal thought experiment, in the past years, an abundance of literature has been devoted to clarify which macroscopic phenomena that show quantum effects can be considered as ``truly macroscopically quantum''. Schr\"odinger's cat has been used as a baseline for comparison to define what a macroscopically quantum system is. After all, the cat is clearly macroscopic and shows a genuinely quantum effect, i.e. being in a superposition of dead and alive. However, characterizing the so-called \textit{macroscopicity} of a quantum state is in general not easy. First, there is not a unique measure that can be applied to all types of quantum states in all scenarios. Some proposed measures are tailored to specific systems like photons or spins, some require that the state to be characterized has the form $\ket{\textrm{dead}} + \ket{\textrm{alive}}$---which is not the case in general, and some others are designed for a given experimental scenario. Second, not all measures focus on the same quantum effect. Even though most of them focus on superposition and entanglement, other effects such as squeezing are also studied. 

In this article, we have introduced some of the proposed measures for macroscopicity by means of two cats A and B that provide two contrasting, visual examples. Our cat A illustrates a true macroscopic quantum state, in contrast to our cat B, which represents a quantum state that, despite being equally large, is not macroscopically quantum, but is instead an accumulation of microscopic quantum effects. Having a concrete visualization of these two cats, where live and dead cells are explicitly depicted, allows us to introduce the different macroscopicity measures in a pedagogical and easy-to-understand way. We have first focused on some relevant measures~\cite{korsbakken2007measurement,sekatski2014size} that are particularly intuitive and then introduced a macroscopicity measure based on the Quantum Fisher Information~\cite{frowis2012measures}, which we have selected due to its broad applicability, both in terms of state types and physical systems. Once the main concepts and intuitions were established with the help of our cats, we discussed more complex, realistic states and their experimental realizations. The explanations and visualizations we provided are complemented by a hands-on activity in which we materialize our quantum cats with cells made out of cardboard. In the activity we propose, students create quantum cats by combining and entangling the cardboard cells, which allows them to analyze in a visual, explicit way whether the cats they create are macroscopically quantum or not. The introduction to macroscopicity measures that we do in Sec.~\ref{sec:macroscopicity_measures} can be directly transferred to this activity.

In addition, we linked the fields of macroscopicity and metrology together and provided examples and visualizations. Macroscopic quantum states are used in metrology as a resource to achieve a higher precision in the estimation of a parameter than other quantum or classical states. Due to this property, the field of metrology has been a source of inspiration for the characterization of macroscopic quantum states. We concluded the manuscript with a summary of state-of-the-art experiments that managed to obtain macroscopic quantum states. Overall, we give the reader a complete and accessible introduction to the field of macroscopicity, with pedagogical, intuitive explanations and visualizations that make this rather advanced field easy to understand for all audiences.

Present experiments involve macroscopic quantum states with $N_{\rm eff}$ of the order of some dozens of particles. While this is fascinating and the progress in this field is fast, it is obvious that we will never have to worry about superpositions between dead and alive cats in a biological sense with $N \propto 10^{26}$, as the noisy  environment will lead to fast decoherence. While Schr\"odingers cat is a beautiful thought experiment, which opened the field of macroscopicity, it is obvious that it should not be taken too literally.

\section{Acknowledgements}

This research was funded in whole or in part by the Austrian Science Fund (FWF) 10.55776/P36009 and 10.55776/P36010. A.L also acknowledges support by the Volkswagen Foundation (Az:97721). For open access purposes, the author has applied a CC BY public copyright license to any author-accepted manuscript version arising from this submission. Finanziert von der Europ\"aischen Union - NextGenerationEU.

\section{Appendix}

\subsection{Quantum Fisher Information for pure states}\label{app:QFI_pure_states}
Quantum states are more generally described by density matrices $\rho = \sum_i \lambda_i \ketbra{i}{i}$, which represent statistical mixtures of (pure) quantum states $\{\ket{i}\}_i$, where each of them occurs with probability $\lambda_i$. Hence, $\sum_i \lambda_i = 1$. When the state is pure, there is only a single element $\ket{\psi}$ in the mixture, thus $\rho = \ketbra{\psi}{\psi}$ ($\lambda_\psi = 1$).

The Quantum Fisher Information (QFI) measures the change of a state $\rho$ towards the infinitesimally closed state $\rho + d\rho$, where the derivative $d\rho = \dot{\rho} d\varphi$ denotes the change with respect to a parameter $\varphi$ that defines a one-dimensional parametrization of the state. Formally, the QFI is defined as
\begin{equation}\label{eq:QFIasBuresMetric}
    \mathcal{F}(\rho, \dot{\rho}) = 2 \sum_{i,j:\lambda_i+\lambda_j\neq 0} \frac{|\langle i |\dot{\rho}| j \rangle|^2}{\lambda_i+\lambda_j},
\end{equation}
where we write $\rho$ as $\rho = \sum_i \lambda_i \ketbra{i}{i}$, as explained above. Typically, an evolution governed by parameter $\varphi$ is written as a parametrization of the state of the form $\rho (\varphi) = e^{-i\varphi D} \rho   e^{i\varphi D}$, where $D$ is an observable that does not depend on $\varphi$. Hence, $\dot{\rho} = i(\rho(\varphi) D - D\rho(\varphi))$.

Taking this parametrization into account, the expression for the QFI reads,
\begin{equation}\label{eq:QFI_densitymatrix}
    \mathcal{F}(\rho, D) = 2 \sum_{i,j:\lambda_i+\lambda_j\neq 0} \frac{(\lambda_i-\lambda_j)^2}{\lambda_i+\lambda_j}|\langle i |D| j \rangle|^2,
\end{equation}
where we have substituted
\begin{align}
    \langle i |\dot{\rho}| j \rangle &= i  \langle i |\rho D| j \rangle - i\langle i |D \rho| j \rangle \nonumber \\
    &= i  \langle i |\sum_k \lambda_k \ketbra{k}{k} D| j \rangle - i  \langle i |D \sum_k \lambda_k \ketbra{k}{k} | j \rangle \nonumber \\
    &= i  \lambda_i \langle i | D| j \rangle - i  \lambda_j \langle i | D| j \rangle = i (\lambda_i-\lambda_j) \langle i | D| j \rangle
\end{align}
in Eq.~\eqref{eq:QFIasBuresMetric}. For pure states $\rho = \ketbra{\psi}{\psi}$, $\lambda_\psi = 1$ and $\lambda_j = 0 \, \forall j\neq \psi$. Therefore, expression~\eqref{eq:QFI_densitymatrix} can be further reduced~\footnote{Note that the sum can be split into case (i) $i=\psi$ and the sum goes over $j$, where $\lambda_j=0$, giving $\sum_j \bra{j}D\ket{\psi}\bra{\psi}D\ket{j} = \sum_j \bra{\psi}D\ket{j}\bra{j}D\ket{\psi}$ and case (ii) $j=\psi$ and the sum goes over $i$, where $\lambda_i=0$, giving $\sum_i \bra{\psi}D\ket{i}\bra{i}D\ket{\psi}$. Both terms are equal, which gives $2 \cdot  \sum_j \bra{\psi}D\ket{j}\bra{j}D\ket{\psi}$.} as follows
\begin{align}
    \mathcal{F}(\ket{\psi}, D) &= 2\cdot 2 \sum_{j} \langle \psi |D| j \rangle \langle j |D|\psi \rangle \nonumber \\
    &= 4 \langle \psi |D^2| \psi \rangle - 4 \langle \psi |D| \psi \rangle \langle \psi |D| \psi \rangle \nonumber \\
    &= 4 (\Delta_{\ket{\psi}} D)^2,
\end{align}
where we have used $\sum_j \ketbra{j}{j} = \id - \ketbra{\psi}{\psi}$. The QFI of a pure state $\ket{\psi}$ is four times the variance of observable $D$ with respect to the state, denoted as $(\Delta_{\ket{\psi}} D)^2$.

\subsection{``Deadness'' variance}\label{app:variances}
In this appendix, we compute the variance of the cat ``deadness'' $D=\sum_{i=1}^N \sigma_z^{(i)}$ for our two example cats, following expression
\begin{equation}
    (\Delta_{\ket{\psi}} D)^2 =  \langle D^2 \rangle_\psi - \langle D \rangle_\psi^2, 
\end{equation}
where $(\Delta_{\ket{\psi}} D)^2$ is the variance of observable $D$ with respect to state $\ket{\psi}$ and $\langle D \rangle_\psi = \langle \psi |D| \psi \rangle$ is the expected value of observable $D$ when the state $\ket{\psi}$ is measured multiple times.

\textbf{(Cat A) GHZ state.---} We start with the first term $\langle \textrm{GHZ} | D^2 | \textrm{GHZ} \rangle$. The expanded expression for $D^2$ in terms of the Pauli matrices reads
\begin{equation}
    D^2=\left( \sum_{i=1}^N \sigma_z^{(i)} \right) \left( \sum_{i=1}^N \sigma_z^{(i)} \right) = \sum_{i=1}^N \id + \sum_{j,k : j\neq k} \sigma_z^{(j)} \sigma_z^{(k)}.
\end{equation}
From this expression, we see that we get $N$ terms of the form $\langle \textrm{GHZ} | \id | \textrm{GHZ} \rangle = 1$, and $(N-1)\cdot N$ terms of the form $\langle \textrm{GHZ} | \sigma_z^{(j)} \sigma_z^{(k)} | \textrm{GHZ} \rangle = 1$, where $j\neq k$. For any pair $j,k$, one gets a computation analogous to
\begin{align}
    \sigma_z^{(1)} \sigma_z^{(2)} | \textrm{GHZ} \rangle &= \frac{1}{\sqrt{2}} ( (+1)^2 \ket{0}^{\otimes N} + (-1)^2 \ket{1}^{\otimes N}) \nonumber \\
    &= | \textrm{GHZ} \rangle,
\end{align}
since $\sigma_z^{(1)} \sigma_z^{(2)} = \sigma_z^{(1)} \otimes \sigma_z^{(2)} \otimes \id \otimes ...\otimes \id$ means that $\sigma_z$ acts on the first and second cell as $\sigma_z \otimes \sigma_z \ket{0} \otimes \ket{0}= (+1)\cdot (+1) \ket{0} \otimes \ket{0}$ and  $\sigma_z \otimes \sigma_z \ket{1} \otimes \ket{1}= (-1)\cdot (-1) \ket{1} \otimes \ket{1}$; and the rest of cells remain unchanged. 

The final expectation value $\langle \textrm{GHZ} | D |\textrm{GHZ}\rangle$ is zero, since there are $N$ terms analogous to
\begin{align}
\langle \textrm{GHZ} | \sigma_z^{(1)} |\textrm{GHZ}\rangle &= \frac{1}{2} (\bra{0}^{\otimes N} + \bra{1}^{\otimes N}) 
 (\ket{0}^{\otimes N} - \ket{1}^{\otimes N}) \nonumber \\
 &= \frac{1}{2} (\langle 00..0|00..0\rangle - \langle 11..1|11..1\rangle) \nonumber \\
 &=\frac{1}{2} (1-1)= 0.
\end{align}

All together, $(\Delta_{\ket{\textrm{GHZ}}} D)^2 = N\cdot 1 + (N-1)N \cdot 1 + 0= N^2.$

\textbf{(Cat B) State $\ket{+}^{\otimes N}$.---} As before, for the first term of the variance one has $N$ terms of the form $\langle +|^{\otimes N}\id \ket{+}^{\otimes N} = 1$. In this case, the $(N-1)\cdot N$ terms of the form $\langle + |^{\otimes N} \sigma_z^{(j)} \sigma_z^{(k)} | + \rangle^{\otimes N}$ cancel each other out. The contribution of the states in the superposition with an even number of live cells cancels out with the contribution from states with an odd number of live cells. As examples of such contributions, we have
\begin{align*}
    \langle 10..0|\sigma_z^{(1)} \sigma_z^{(2)} | 10..0\rangle &= (-1)\cdot (+1) \langle 10..0| 10..0\rangle = -1 \nonumber \\
    \langle 11..0|\sigma_z^{(1)} \sigma_z^{(2)} | 11..0\rangle &= (-1)\cdot (-1) \langle 11..0| 11..0\rangle = +1.
\end{align*}

Analogously, the expectation value $\bra{+}^{\otimes N} D \ket{+}^{\otimes N}$ is zero, since its terms cancel each other out. Examples of contributions in this case are
\begin{align*}
     \langle 00..0|\sum_{i=1}^N \sigma_z^{(i)} | 00..0\rangle &= (+1+..+1) \langle 00..0| 00..0\rangle ) = +N \nonumber \\
     \langle 11..1|\sum_{i=1}^N \sigma_z^{(i)} | 11..1\rangle &= (-1-..-1) \langle 11..1| 11..1\rangle ) = -N.
\end{align*}

All together, $(\Delta_{\ket{+}^{\otimes N}} D)^2 = N\cdot 1 + 0 + 0= N.$

% The expectation value $\langle \textrm{GHZ} | D |\textrm{GHZ}\rangle$ is zero, since 
% $$(\langle 0 | + \langle 1 |) \sigma_z  (| 0 \rangle + | 1 \rangle) = (\langle 0 + \langle 1 |) (| 0 \rangle - | 1 \rangle )= 0$$
% %
% Concerning $\langle \textrm{GHZ} | D^2 | \textrm{GHZ} \rangle$, we obtain $N^2$ summands

% \bea
% \langle D^2\rangle  = N \bf{1} + \langle \sum_{k \neq  l} \sigma^{(k)}_z \sigma^{(l)}_z\rangle
% \eea
% %
% For the GHZ-state, due to $(\pm 1)^2 = 1$, all these summands add up to $N^2$, leading to $(\Delta_{\ket{\textrm{GHZ}}} D)^2  = N^2$.

\subsection{Dicke state}
\label{Dicke}

Calculate $N_{\rm eff}$ for the generalized $W$ state

\bea \label{Ws}
|W_N\rangle = \frac{1}{\sqrt{N}} (|1 0 0, ..., 0 \rangle + |0 1 0, ... 0 \rangle + 
\\ \nonumber
|0 0 1, ... 0 \rangle + ... + |0 0 0, ... 1 \rangle
\eea
Note that this is the Dicke-state $| D_N^{(1)}\rangle$. As local operator, we choose the spin-flip operator $\sigma_x$, as this turn out to lead to maximal variance. Let $\sigma^{(k)}_x$ act on  the $k-$th qubit via interchange $|0_k \rangle \leftrightarrow |1_k \rangle$. 
\begin{figure}
\begin{center}
\includegraphics[scale=0.99]{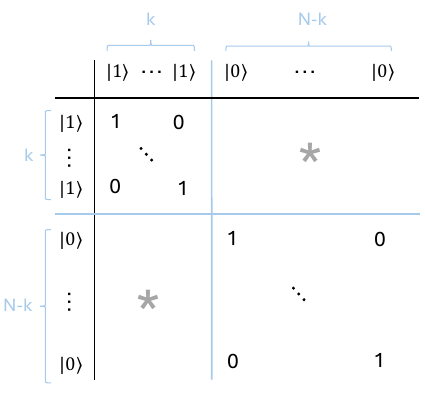}
\end{center}
\caption{Non-zero contributions to the variance $\langle A^2\rangle$ for $A=\sum_k \sigma_x^{(k)}$ come from the $N$ diagonal terms and from the $2 k (N-k)$ off-diagonal parts where the operator $\sigma_x^r \sigma_x^s$ interchanges $|0_p 1_q\rangle$ to $|0_q 1_p\rangle$  } \label{DickeV}
\end{figure}
First, we evaluate this expression for the generalized $W$ state, then we proceed to the general Dicke-state $| D_N^{(k)}\rangle$.
The first summand contains $N^2$ operators of the form
\bea
\langle A^2\rangle  = N \bf{1} + \langle \sum_{k \neq  l} \sigma^{(k)}_x \sigma^{(l)}_x\rangle
\eea
Consider the off-diagonal term 
$\langle W_N | \sigma_x^{(k)} \sigma_x^{(l)} |W_N\rangle$, where the state of the k-th and l-th qubit are flipped. If the original states were $0_k 0_l$, then the resulting flipped state $1_k 1_l$ would be orthogonal to $| W_N \rangle$ and the result would vanish. Non-zero contributions arise only for $1_k 0_l$ or  $0_k 1_l$.

There are N possible locations for the state $1$, see equation \ref{Ws}. This corresponds to the number of terms in the $|W_N\rangle$-state and in turn to the normalization $1/N$. Without loss of generality, we can thus skip the normalization and consider just the term $| 1, 0, ..., 0 \rangle$ and count the number of non-vanishing contributions in $\sigma^{(1)}_x \sigma^{(k)}$ and  $\sigma^{(k)}_x \sigma^{(1)}$ for $k\neq 1$. Obviously, there are $2 (N-1)$ such contributions, leading to the result
\bea\label{VarW}
\langle W_N | A^2 |W_N\rangle  = N  +  2 (N-1)
\eea
As $\langle W_N | A |W_N\rangle$ is zero since any single spin-flip is orthogonal to $|W_N\rangle$, the expression (\ref{VarW}) coincides with the variance.

A similar argument holds for the general Dicke-state $| D_N^{(k)}\rangle$, which is a superposition of all $N!/(k! (N-k)!)$ possible combinations of terms containing $k$-times the state $0$ and $(N-k)$-times the state $0$. Again, we can consider the state $|1111_k 000000_N\rangle$ and discuss its contribution under the action of the spin-flip operator $\sigma^{(p)}_x \sigma^{(q}$. Due to the summation $p \neq q$, the result we obtain for this term will occur $N!/(k! (N-k)!)$-times, which cancels the normalization. Thus, it is sufficient just to consider this contribution, as shown in Fig. \ref{DickeV}. Non-zero contributions can only arise under interchange of $1 \rightarrow 0$, which means that the combinations  $p \in \{1, k\}$, $q \in \{k+1, N\}$ and  $p \in \{k+1, N\}$, $q \in \{1, k\}$ are relevant. Thus, we find
\bea\label{VarD}
\langle D_N^{(k)} | A^2 |D_N^{(k)} \rangle  = N  +  2 k (N-k)
\eea

The measure for macroscopicity thus leads to
\bea\
N_{\rm eff} = \frac{1}{N} \langle D_N^{(k)} | ( \Delta A)^2 |D_N^{(k)} \rangle  = 1  +  2 \frac{k (N-k)}{N}
\eea
For $k=1$, the state is $not$ macroscopic as $N_{\rm eff}$ is of order $o(1)$, wile for $k=N/2$, the state $is$ macroscopic as  $N_{\rm eff}$ is of order $N$. The calculation shows that the difference can be traced back to number of possible spin-flip-partners available, as visualized in Fig. \ref{DickeV}.

\subsection{Cram\'er-Rao bound}\label{app:cramer-rao_bound}
In this appendix, we compute the Cram\'er-Rao bound for the example of parameter estimation we introduced in Section~\ref{sec:macrostates_and_metrology}. We compute the bound using Eq.~\eqref{eq:FI_metrology} for each of our two example states. In both cases, we consider that (i) the probability $p_1(\varphi)$ is the probability that we obtain the initial state when we measure the state after the evolution, i.e.,
\begin{equation}
    p_1(\varphi) = |\langle \psi_{\textrm{init}} | \psi_{\textrm{fin}}\rangle |^2,
\end{equation}
and (ii) $p_2(\varphi)= 1-p_1(\varphi)$.

\textbf{(A) GHZ state.---} In this case, $|\psi_{\textrm{init}}\rangle = |\textrm{GHZ}\rangle$, and $|\psi_{\textrm{fin}}\rangle = U^{\otimes N} \ket{\textrm{GHZ}}$ (see Eq.~\eqref{eq:evolved_GHZ_metrology}). Hence,
\begin{align}
    p_1(\varphi) &= \left|\frac{1}{2} (1+e^{iN\varphi)} \right|^2 = \cos^2{\frac{N\varphi}{2}}, \nonumber \\
    p_2(\varphi) &= \sin^2{\frac{N\varphi}{2}}.
\end{align}
The derivative of the logarithm of these expressions thus reads,
\begin{align}
    \frac{d}{d\varphi} \log{p_1(\varphi)} &= -N\cdot \frac{\sin{(N\varphi /2)}}{\cos{(N\varphi/2)}}, \nonumber \\
    \frac{d}{d\varphi} \log{p_2(\varphi)} &= N\cdot \frac{\cos{(N\varphi /2)}}{\sin{(N\varphi/2)}}.
\end{align}
Substituting the above expressions in Eq.~\eqref{eq:FI_metrology}, one gets 
\begin{equation}
    \mathcal{F}(\varphi) =  N^2\cdot \sin^2{\frac{N\varphi}{2}} +  N^2 \cdot \cos^2{\frac{N\varphi}{2}}= N^2,
\end{equation}
which gives a Cram\'er-Rao bound $\Delta \varphi \geq 1/N$.

\textbf{(B) State $\ket{+}^{\otimes N}$.---} Since the $N$ qubits are independent, we consider here a single-qubit evolution, which is then repeated $N$ times. Therefore, $|\psi_{\textrm{init}} \rangle= |+\rangle$, and $|\psi_{\textrm{fin}} \rangle= U \ket{+} = 1/\sqrt{2} ( \ket{0} + e^{i\varphi} \ket{1})$. The expressions for $p_1(\varphi)$, $p_2(\varphi)$ and their derivatives are analogous to the previous case with the GHZ state. The final expression for the QFI reads,
\begin{equation}
    \mathcal{F}(\varphi) = \cos^2{\frac{\varphi}{2}}\cdot \frac{\sin^2{(\varphi /2)}}{\cos^2{(\varphi/2)}} + \sin^2{\frac{\varphi}{2}} \cdot \frac{\cos^2{(\varphi /2)}}{\sin^2{(\varphi/2)}} = 1.
\end{equation}
In this case, the Cram\'er-Rao bound reads $\Delta \varphi \geq \frac{1}{\sqrt{N}\cdot \sqrt{\mathcal{F}(\varphi)}} = 1/\sqrt{N}$, where the factor $1/\sqrt{N}$ is the precision enhancement introduced by the $N$ repetitions of the same process.

We see how the macroscopic quantum state $\ket{\textrm{GHZ}}$ consisting of $N$ entangled qubits gives an enhancement in precision compared to the case where $N$ independent qubits undergo the sensing process.

\bibliographystyle{unsrt}
\bibliography{biblio}

\begin{thebibliography}{10}

\bibitem{frowis2012measures}
Florian Fr{\"o}wis and Wolfgang D{\"u}r.
\newblock Measures of macroscopicity for quantum spin systems.
\newblock {\em New Journal of Physics}, 14(9):093039, 2012.

\bibitem{korsbakken2007measurement}
Jan~Ivar Korsbakken, K~Birgitta Whaley, Jonathan Dubois, and J~Ignacio Cirac.
\newblock Measurement-based measure of the size of macroscopic quantum
  superpositions.
\newblock {\em Physical Review A}, 75(4):042106, 2007.

\bibitem{sekatski2014size}
Pavel Sekatski, Nicolas Sangouard, and Nicolas Gisin.
\newblock Size of quantum superpositions as measured with classical detectors.
\newblock {\em Physical Review A}, 89(1):012116, 2014.

\bibitem{topology}
S.~Heusler, P.~Schlummer, and M.S. Ubben.
\newblock The topological origin of quantum randomness.
\newblock {\em Symmetry}, 13(4), 2021.

\bibitem{Leggett1980}
A.~J. Leggett.
\newblock {Macroscopic Quantum Systems and the Quantum Theory of Measurement}.
\newblock {\em Progress of Theoretical Physics Supplement}, 69:80--100, 03
  1980.

\bibitem{Frowis2017}
Florian Fr\"owis, Pavel Sekatski, Wolfgang D\"ur, Nicolas Gisin, and Nicolas
  Sangouard.
\newblock Macroscopic quantum states: Measures, fragility, and implementations.
\newblock {\em Rev. Mod. Phys.}, 90:025004, May 2018.

\bibitem{dur2014visualization}
Wolfgang D{\"u}r and Stefan Heusler.
\newblock Visualization of the invisible: The qubit as key to quantum physics.
\newblock {\em The Physics Teacher}, 52(8):489--492, 2014.

\bibitem{giovannetti2004quantum}
Vittorio Giovannetti, Seth Lloyd, and Lorenzo Maccone.
\newblock Quantum-enhanced measurements: beating the standard quantum limit.
\newblock {\em Science}, 306(5700):1330--1336, 2004.

\bibitem{Deco2018}
Stefan Heusler and Wolfgang D\"ur.
\newblock Modeling decoherence with qubits.
\newblock {\em Eur. J. Phys.}, 39:025406, 2018.

\bibitem{monz2011fourteenqubit}
Thomas Monz, Philipp Schindler, Julio~T Barreiro, Michael Chwalla, Daniel Nigg,
  William~A Coish, Maximilian Harlander, Wolfgang H{\"a}nsel, Markus Hennrich,
  and Rainer Blatt.
\newblock 14-qubit entanglement: Creation and coherence.
\newblock {\em Physical Review Letters}, 106(13):130506, 2011.

\bibitem{hosten2016measurement}
Onur Hosten, Nils~J Engelsen, Rajiv Krishnakumar, and Mark~A Kasevich.
\newblock Measurement noise 100 times lower than the quantum-projection limit
  using entangled atoms.
\newblock {\em Nature}, 529(7587):505--508, 2016.

\bibitem{marciniak2022optimal}
Christian~D Marciniak, Thomas Feldker, Ivan Pogorelov, Raphael Kaubruegger,
  Denis~V Vasilyev, Rick van Bijnen, Philipp Schindler, Peter Zoller, Rainer
  Blatt, and Thomas Monz.
\newblock Optimal metrology with programmable quantum sensors.
\newblock {\em Nature}, 603(7902):604--609, 2022.

\bibitem{vahlbruch2016detection}
Henning Vahlbruch, Moritz Mehmet, Karsten Danzmann, and Roman Schnabel.
\newblock Detection of 15 db squeezed states of light and their application for
  the absolute calibration of photoelectric quantum efficiency.
\newblock {\em Physical review letters}, 117(11):110801, 2016.

\bibitem{deleglise2008reconstruction}
Samuel Deleglise, Igor Dotsenko, Clement Sayrin, Julien Bernu, Michel Brune,
  Jean-Michel Raimond, and Serge Haroche.
\newblock Reconstruction of non-classical cavity field states with snapshots of
  their decoherence.
\newblock {\em Nature}, 455(7212):510--514, 2008.

\bibitem{vlastakis2013deterministically}
Brian Vlastakis, Gerhard Kirchmair, Zaki Leghtas, Simon~E Nigg, Luigi Frunzio,
  Steven~M Girvin, Mazyar Mirrahimi, Michel~H Devoret, and Robert~J Schoelkopf.
\newblock Deterministically encoding quantum information using 100-photon
  schr{\"o}dinger cat states.
\newblock {\em Science}, 342(6158):607--610, 2013.

\bibitem{wang2016schrodinger}
Chen Wang, Yvonne~Y Gao, Philip Reinhold, Reinier~W Heeres, Nissim Ofek, Kevin
  Chou, Christopher Axline, Matthew Reagor, Jacob Blumoff, KM~Sliwa, et~al.
\newblock A schr{\"o}dinger cat living in two boxes.
\newblock {\em Science}, 352(6289):1087--1091, 2016.

\bibitem{kienzler2016observation}
Daniel Kienzler, C~Fl{\"u}hmann, Vlad Negnevitsky, H-Y Lo, M~Marinelli,
  D~Nadlinger, and Jonathan~P Home.
\newblock Observation of quantum interference between separated mechanical
  oscillator wave packets.
\newblock {\em Physical review letters}, 116(14):140402, 2016.

\bibitem{kovachy2015quantum}
Tim Kovachy, Peter Asenbaum, Chris Overstreet, Christine~A Donnelly, Susannah~M
  Dickerson, Alex Sugarbaker, Jason~M Hogan, and Mark~A Kasevich.
\newblock Quantum superposition at the half-metre scale.
\newblock {\em Nature}, 528(7583):530--533, 2015.

\bibitem{aspelmeyer2014cavity}
Markus Aspelmeyer, Tobias~J Kippenberg, and Florian Marquardt.
\newblock Cavity optomechanics.
\newblock {\em Reviews of Modern Physics}, 86(4):1391--1452, 2014.

\bibitem{barzanjeh2022optomechanics}
Shabir Barzanjeh, Andr{\'e} Xuereb, Simon Gr{\"o}blacher, Mauro Paternostro,
  Cindy~A Regal, and Eva~M Weig.
\newblock Optomechanics for quantum technologies.
\newblock {\em Nature Physics}, 18(1):15--24, 2022.

\bibitem{gerlich2011quantum}
Stefan Gerlich, Sandra Eibenberger, Mathias Tomandl, Stefan Nimmrichter, Klaus
  Hornberger, Paul~J Fagan, Jens T{\"u}xen, Marcel Mayor, and Markus Arndt.
\newblock Quantum interference of large organic molecules.
\newblock {\em Nature communications}, 2(1):263, 2011.

\bibitem{Note1}
Note that the sum can be split into case (i) $i=\psi $ and the sum goes over
  $j$, where $\lambda _j=0$, giving $\DOTSB \sum@ \slimits@ _j \left \langle j
  \right |D\left | \psi \right \rangle \left \langle \psi \right |D\left | j
  \right \rangle = \DOTSB \sum@ \slimits@ _j \left \langle \psi \right |D\left
  | j \right \rangle \left \langle j \right |D\left | \psi \right \rangle $ and
  case (ii) $j=\psi $ and the sum goes over $i$, where $\lambda _i=0$, giving
  $\DOTSB \sum@ \slimits@ _i \left \langle \psi \right |D\left | i \right
  \rangle \left \langle i \right |D\left | \psi \right \rangle $. Both terms
  are equal, which gives $2 \cdot \DOTSB \sum@ \slimits@ _j \left \langle \psi
  \right |D\left | j \right \rangle \left \langle j \right |D\left | \psi
  \right \rangle $.

\end{thebibliography}

\end{document}